\documentclass[reprint,aps,prb,amsmath,superscriptaddress,bibnotes,longbibliography]{revtex4-2}

\usepackage{hyperref}   
\hypersetup{
	colorlinks=true,    
	linkcolor=blue,     
	citecolor=blue,     
	urlcolor=blue       
}
\usepackage{mathptmx,newtxtext,newtxmath,xspace}
\usepackage{amsbsy,bm,bbold}
\usepackage{graphicx,color,xcolor}
\usepackage{fancyhdr}

\usepackage{psfrag}
\usepackage{enumitem}

\DeclareTextCompositeCommand{\r}{OT1}{A}{%
	\leavevmode\vbox{%
		\offinterlineskip
		\ialign{\hfil##\hfil\cr\char23\cr\noalign{\kern-1.15ex}A\cr}%
	}%
}
\usepackage{array}
\usepackage{longtable}
\usepackage{geometry}

\begin{document}
	
	\widetext
	\date{\today}
	
	\title{Nature of the ferromagnet-paramagnet transition in Y$_{1-x}$Ca$_{x}$TiO$_{3}$}    

	\author{S. Hameed}
	\affiliation{School of Physics and Astronomy, University of Minnesota, Minneapolis, Minnesota 55455, U.S.A.}
	\affiliation{Max-Planck-Institute for Solid State Research, Heisenbergstraße 1, 70569 Stuttgart, Germany}
	\author{I. Khayr}
	\affiliation{School of Physics and Astronomy, University of Minnesota, Minneapolis, Minnesota 55455, U.S.A.}
	\author{J. Joe}
	\affiliation{School of Physics and Astronomy, University of Minnesota, Minneapolis, Minnesota 55455, U.S.A.}
	\author{G. Q. Zhao}
	\affiliation{Department of Physics, Columbia University, New York, New York 10027, USA.}
	\affiliation{Institute of Physics, Chinese Academy of Sciences, Beijing 100190, China.}
	\author{Y. Cai}
	\affiliation{TRIUMF, Vancouver, British Columbia, V6T 2A3, Canada.}
	\affiliation{Department of Physics, Columbia University, New York, New York 10027, USA.}
	\author{K. M. Kojima }
	\affiliation{TRIUMF, Vancouver, British Columbia, V6T 2A3, Canada.}
	\author{S. Chi}
	\affiliation{Neutron Scattering Division, Oak Ridge National Laboratory, Oak Ridge, Tennessee 37831, USA}
	\author{T. J. Williams}
	\affiliation{Neutron Scattering Division, Oak Ridge National Laboratory, Oak Ridge, Tennessee 37831, USA}
	\author{M. Matsuda}
	\affiliation{Neutron Scattering Division, Oak Ridge National Laboratory, Oak Ridge, Tennessee 37831, USA}
	\author{Y. J. Uemura}
	\affiliation{Department of Physics, Columbia University, New York, New York 10027, USA.}	
	\author{M. Greven}
	\affiliation{School of Physics and Astronomy, University of Minnesota, Minneapolis, Minnesota 55455, U.S.A.}
	\begin{abstract}
 	Neutron scattering, magnetometry, and muon spin rotation ($\mu$SR) measurements were performed to investigate the magnetic order and spin dynamics across the ferromagnet-to-paramagnet transition in the hole-doped Mott insulator Y$_{1-x}$Ca$_x$TiO$_3$. We find that the transition proceeds through a volume-wise phase separation into ferromagnetic and paramagnetic regions. Spin fluctuations with a characteristic timescale of $\sim$ 0.1 $\mu$s, as detected via $\mu$SR, are observed to appear at Ca concentrations $x \geq 0.10$. The magnetic phase separation, accompanied by a modest dynamic response, represents a novel behavior in Mott systems near the loss of magnetic order. It is linked to a previously observed insulator-metal transition and the associated electronic phase separation into hole-poor Mott insulating and hole-rich metallic phases for $0 < x < 0.50$. In particular, the $x$-dependence of the paramagnetic volume fraction strongly correlates with that of the volume fraction of the hole-rich metallic phase. The spin-wave spectra reveal a doping-induced crossover from isotropic to two-dimensional anisotropic exchange interactions, reflecting substantial changes in the orbital state with increasing Ca content.
 	
	\end{abstract}

	\pacs{}
	\maketitle
 	\section{Introduction}
	
	Transition-metal oxides with orbital degeneracy have garnered significant interest due to the diverse spin-orbital ground states they exhibit, driven by competing mechanisms that lift the orbital degeneracy \cite{Goodenough1955,Kanamori1959,Kugel1982,Tokura2000,Mochizuki2004,Khaliullin2005,Radhakrishnan2024}. The rare-earth titanates (RETiO$_3$, where RE represents a rare-earth element) are Mott insulators, with Ti ions in the +3 valence state, hosting a single $d$ electron in their $t_{2g}$ orbitals. The size mismatch between RE and Ti induces a GdFeO$_3$-type distorted perovskite structure, characterized by a Ti-O-Ti bond angle that significantly deviates from 180°. This bond angle is a crucial determinant of the system's magnetic ground state: compounds with smaller bond angles exhibit ferromagnetic (FM) order, whereas those with larger angles display antiferromagnetic (AF) order \cite{MacLean1979,Greedan1985,Mochizuki2004,Zhou2005}.
	
	Two competing theoretical frameworks have attempted to explain the FM order observed in materials with smaller rare-earth ionic radii (and hence, smaller Ti-O-Ti bond angle), such as YTiO$_3$ and GdTiO$_3$. The first is a crystal-field-based model, which suggests that a $d$-type Jahn-Teller distortion lifts the orbital degeneracy and stabilizes a nearly antiferro-type orbital order, leading to FM behavior \cite{Mochizuki2000,Mochizuki2001}. However, spin-wave spectra for YTiO$_3$ reveal three-dimensional isotropic behavior, inconsistent with the highly anisotropic orbital order predicted by this model \cite{Ulrich2002,HameedYLaSW2023}. An alternative explanation involves spin-orbital superexchange interactions and assumes that the crystal field does not fully lift the orbital degeneracy. This scenario proposes a weak orbital order that retains the cubic symmetry of the spin-exchange bonds, accompanied by strong orbital fluctuations, which accounts for both the FM ground state and the observed isotropic spin-wave spectrum \cite{Khaliullin2002,Khaliullin2003,Khaliullin2005}. Both models explain the transition from FM to AF order with increasing Ti-O-Ti bond angle and attribute this change to the role of $t_{2g}-e_g$ hybridization in stabilizing the FM order \cite{Khaliullin2003,Khaliullin2005}. 
	
	A recent electron spin resonance (ESR) study revealed a small orbital gap of approximately 10 meV in YTiO$_3$ \cite{Najev2024}. The gap was found to decrease with increasing Ca or La substitution, both of which lower the Curie temperature ($T_C$) and suppress the FM order. This finding suggests that the complete lifting of orbital degeneracy is closely linked to the stabilization of the FM phase, consistent with the crystal-field picture. Furthermore, orbital fluctuations are expected to increase with Ca and La substitution, due to the decrease of the orbital gap \cite{Najev2024}. Besides the transition to a paramagnetic (PM) state, Ca doping also induces an insulator-metal transition. The Ca-doped system exhibits electronic phase separation into hole-poor Mott-insulating and hole-rich metallic regions for $0 < x < 0.50$, with the percolation of the metallic regions leading to metallic transport for $x > 0.30$ \cite{HameedYCa2021}.
	
	Motivated by these observations, we investigate the magnetic order and spin dynamics in Y$_{1-x}$Ca$_x$TiO$_3$ via neutron scattering, magnetometry, and muon spin rotation ($\mu$SR). We find that the FM-to-PM transition in Y$_{1-x}$Ca$_x$TiO$_3$ proceeds through a volume-wise phase separation into FM and PM regions. The PM volume fraction exhibits an $x$-dependence that strongly correlates with that of the volume fraction of the hole-rich metallic phase, suggesting that the PM regions correspond to the hole-rich metallic phase. Spin fluctuations with a characteristic timescale of $\sim$ 0.1 $\mu$s, as detected via $\mu$SR, appear for $x \geq 0.10$. The emergence of magnetic phase separation with a modest dynamic response indicates a unique behavior in Mott systems nearing the disappearance of static magnetism, driven by the underlying insulator-metal transition and the accompanying electronic phase separation. Additionally, spin-wave measurements show an increasing $ab$-plane $vs.$ $c$-axis spin-exchange anisotropy with increasing Ca doping, indicative of substantial changes in the orbital state.

	\section{Experimental methods}

	Single crystals of Y$_{1-x}$Ca$_{x}$TiO$_{3}$ were melt-grown with the optical traveling solvent floating zone (TSFZ) technique and well-characterized post-growth, as described previously~\cite{HameedGrowth2021}. 
	
	DC magnetic susceptibility measurements were carried out using a Quantum Design, Inc. Magnetic Property Measurement System (MPMS). AC susceptibility measurements were performed with the MPMS using a custom-built probe consisting of two matched detection coils, with the excitation field supplied by the integrated coil. We used an excitation frequency of 3.63 kHz. The samples were pre-aligned using a Laue diffractometer, with the magnetic field applied along the $c$-axis. 
	
	$\mu$SR measurements were performed on the M15 and M20 surface muon beam lines with the DR spectrometer and LAMPF spectrometer, respectively, at the Centre for Molecular and Materials Science at TRIUMF in Vancouver, Canada. The initial muon spin was antiparallel to its momentum, and the sample [001] axis was orientated approximately perpendicular to the muon spin direction. We used samples with typical mass $\sim$ 1 g. A dilution refrigerator at M15 and a helium gas-flow cryostat at M20 were used for temperature control down to 20 mK and 2 K, respectively. The data were analyzed using musrfit \cite{Suter2012}.

	 Triple-axis neutron spectroscopy was performed with the HB-1 and HB-3 thermal-neutron triple-axis spectrometers at the High Flux Isotope Reactor, located at Oak Ridge National Laboratory. We studied $x = 0$ and $x = 0.05$ crystals with masses $\sim3$ g, and an $x = 0.15$ sample composed of 20 co-aligned crystals with a total mass of $\sim$ 12 g. The larger sample mass for $x = 0.15$ was needed to detect the weaker inelastic signal due to the reduced ordered magnetic moment and strongly damped spin waves \cite{HameedGrowth2021}. The detailed procedure adopted to ensure chemical homogeneity of the samples can be found in Ref.~\cite{HameedGrowth2021}. For the inelastic neutron scattering measurements, the energies of the scattered neutrons were fixed at either 14.7 or 30.5 meV. For neutron diffraction measurements, a fixed final energy of 13.5 meV was used. We used collimations of $48'-80'- $sample$ - 80'- 120'$, and PG filters after the sample to eliminate higher-order neutrons. A $^4$He-cryostat was used to reach temperatures down to 2 K. The samples were mounted in the $(0KL)$ scattering plane, and the spin-wave excitations were characterized via energy and momentum scans; each scan was performed twice: at 2 K and at 25 K, well below and above the magnetic transition temperatures, respectively.

	\section{Results}
	
	\subsection{Magnetometry and neutron diffraction}
	\begin{figure}
		\includegraphics[width=0.48\textwidth]{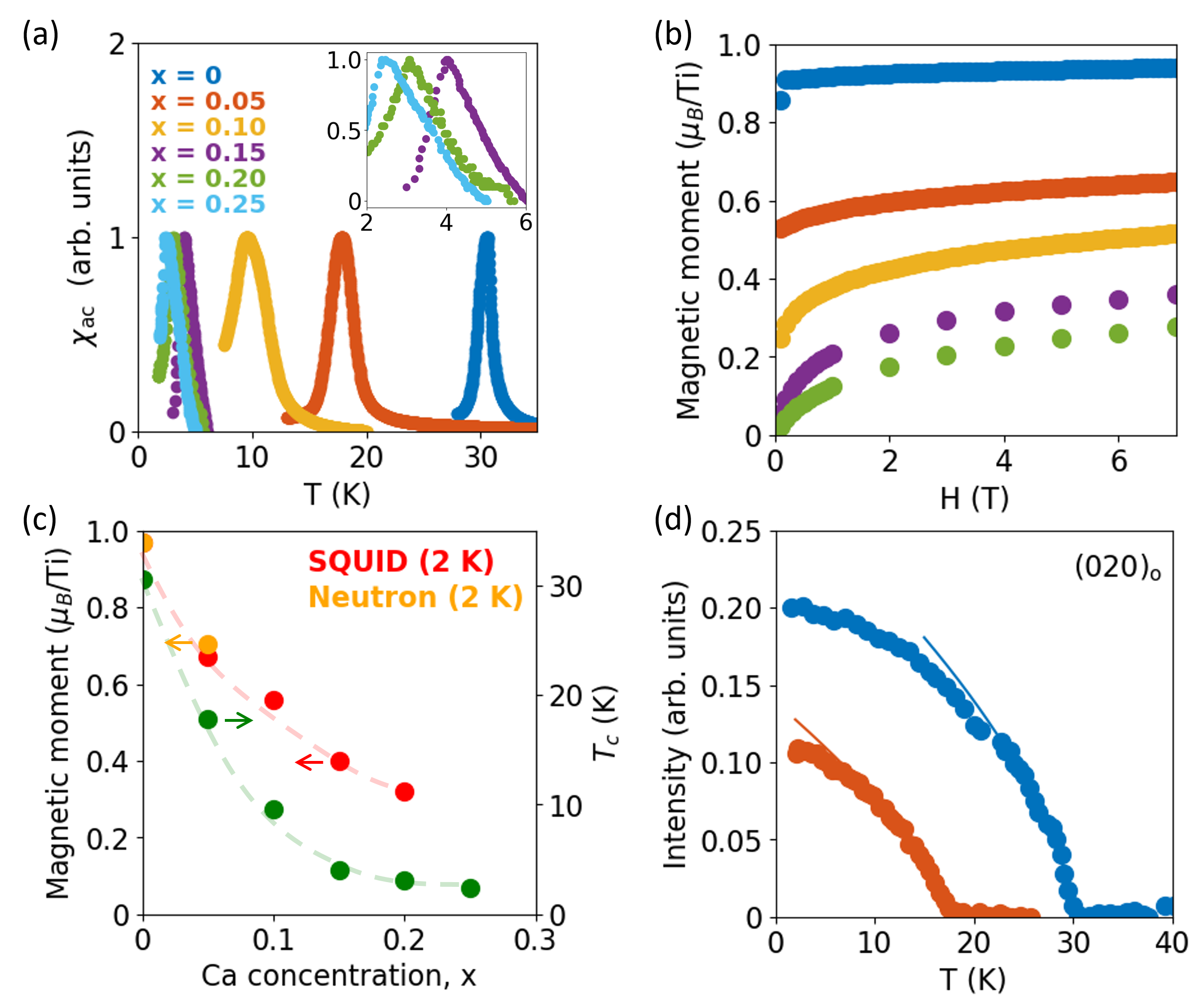}
		\caption{(a) Doping and temperature dependence of the AC magnetic susceptibility. The susceptibility in the PM state at the highest measured temperatures was subtracted, and the resultant data were normalized to the peak value at $T_C$. The inset shows a magnified view of the data for $x > 0.10$. (b) Magnetic-field dependence of the magnetic moment per Ti ion along the $c$-axis at $T$ = 2 K, measured using SQUID magnetometry. The samples were zero-field cooled and measured in an ascending field. (c) Ca-doping dependences of $T_C$, estimated from the peak positions in (a) (right axis), and the saturation FM moment, estimated from an exponential extrapolation of the magnetometry data above 2 T in (b) (left axis). (d) Temperature dependences of the $(020)_{{\text{o}}}$ FM Bragg intensity obtained from neutron diffraction for $x = 0$ and 0.05. The lines are power-law fits to the order-parameter form $I \propto (1-T/T_{C})^{2\beta}$, as described in the text. The square root of the measured $(020)_\text{o}$ neutron diffraction intensity at 2 K, normalized to match the magnetic moment obtained from SQUID magnetometry at $x = 0$ is included in (c) for comparison.}
		\label{fig:SQUID_neutron}
	\end{figure}	
	
	\begin{figure}
		\includegraphics[width=0.48\textwidth]{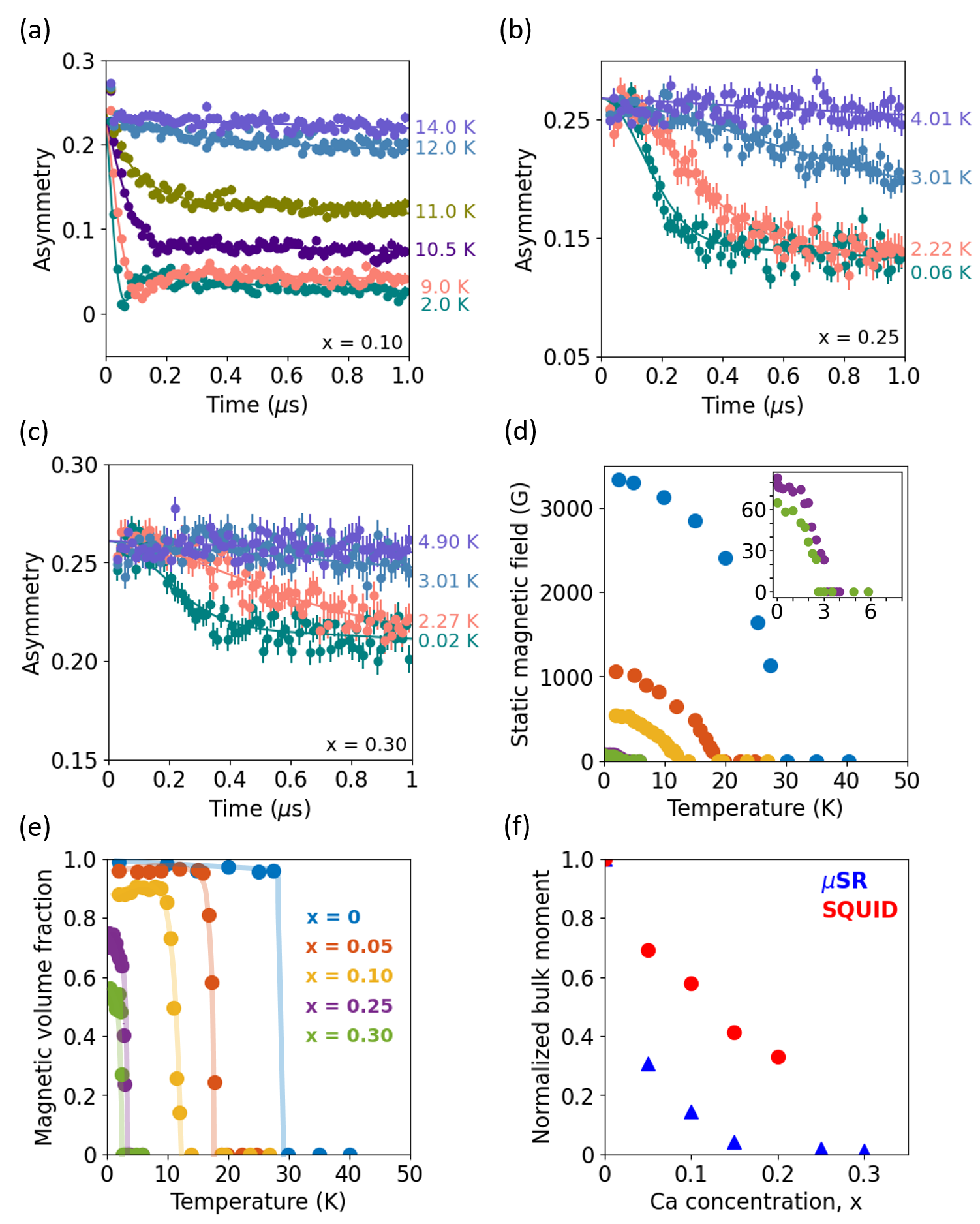}
		\caption{(a-d) ZF $\mu$SR spectra at different temperatures for (a) $x = 0.10$, (b) $x = 0.25$, and (c) $x = 0.30$. Solid lines are fits to the data (see text). (d) Static local field (e) and magnetic volume fraction $vs.$ temperature for different Ca concentrations, obtained from fits to the ZF spectra. The data for $x = 0$ in (d,e) are reproduced from Ref.~\cite{HameedYLa2021}. The inset in (d) shows an enlarged view of the $x = 0.25$ and 0.30 data. (f) $\mu$SR measure of the bulk ordered moment (local moment multiplied by magnetic volume fraction), compared with the extrapolated bulk FM moment from Fig.~\ref{fig:SQUID_neutron}(c). The data are normalized by the respective $x = 0$ values.}
		\label{fig:ZFMuSR}
	\end{figure}
	
	Figure.~\ref{fig:SQUID_neutron}(a) displays the temperature dependence of the AC magnetic susceptibility $\chi_{ac}$ obtained for Y$_{1-x}$Ca$_{x}$TiO$_{3}$ with $x = 0-0.25$. The Ca concentration dependence of the Curie temperature $T_C$, defined here as the peak position of $\chi_{\text{ac}}$, is shown on the right axis in Fig.~\ref{fig:SQUID_neutron}(c). A strong suppression of $T_C$ is observed up to $x = 0.15$, beyond which the rate of decrease of the transition temperature with increasing Ca concentration appears to slow. This result is consistent with a recent uniaxial strain study of a $x = 0.15$ sample, where the magnetic signal was observed to disappear while $T_C$ remained unaffected by strain \cite{Najev2022}. These observations are consistent with a first-order nature of the Ca-doping-induced FM-PM transition, as we return to in Section~\ref{MuSR}. Figure~\ref{fig:SQUID_neutron}(b) shows the magnetic field dependence of the magnetic moment at 2 K for a range of Ca doping levels $x = 0-0.20$. The samples were zero-field cooled, and the magnetic field was applied along the $c$-axis. Upon doping, the high-field magnetization decreases rapidly. We use an exponential extrapolation  of the magnetic moments above 2 T, which is known to be applicable to soft ferromagnetic materials \cite{Umenei2011,HameedYLa2021}, to extract the saturation magnetic moment. As shown on the left axis of Fig.~\ref{fig:SQUID_neutron}(c), a strong drop in the saturation magnetization is observed, from $\sim$ 0.97 $\mu$$_{\text{B}}$/Ti for $x = 0$ to $\sim$ 0.67 $\mu$$_{\text{B}}$/Ti for $x = 0.05$ and further to $\sim$ 0.32 $\mu$$_{\text{B}}$/Ti for $x = 0.20$. Note that we have not considered a potential PM contribution to the field dependence of the magnetic moments in this analysis, although we discuss this below. Such a PM component could also influence the exponential form of the high-field magnetization, as described by the Langevin theory of paramagnetism. For $x = 0.15$ and 0.20, 2 K is close to the respective $T_{C}$, which explains the absence of a strong initial increase in the magnetization at low applied fields — an increase typically expected from the reorientation of FM domains under a magnetic field, as observed for $x = 0$ to 0.10.
	
	Figure~\ref{fig:SQUID_neutron}(d) shows the temperature dependence of the $(020)_{{\text{o}}}$ FM Bragg reflection obtained from neutron diffraction for two Ca doping levels, $x = 0$ and $x = 0.05$. A relatively weak temperature-independent nuclear contribution was subtracted. In order to obtain the relative strength of the elastic magnetic response for the two samples, the net intensity was normalized by the nuclear contribution to the $(020)_{{\text{o}}}$ Bragg reflection, obtained in the PM state at $T > T_C$. Power-law fits of the form $I \propto (1-T/T_{C})^{2\beta}$ in the vicinity of $T_{C}$ yield critical exponent estimates $\beta = 0.32(3)$ ($x = 0$) and $\beta = 0.34(2)$ ($x = 0.05$), consistent with the 3D Ising value of 0.3265(7) \cite{Odor2004}, as expected due to the system's 3D Ising anisotropy \cite{HameedYLa2021}. The relative strength of the square root of the intensity at 2 K, which serves as a measure of the FM magnetic moment, is shown in Fig.~\ref{fig:SQUID_neutron}(c) and found to be in good agreement with the magnetization results.

	\subsection{Muon spin rotation}
	\label{MuSR}
	$\mu$SR is a highly sensitive local probe of magnetism that is capable of detecting both static and dynamic magnetism. Positively charged spin-polarized muons get implanted at interstitial sites and undergo Larmor precession due to magnetic fields present in the sample. Detection of the asymmetric decay of the muon enables independent measurements of both the local magnetic moment and the magnetically ordered volume fraction. This technique is therefore highly complementary to magnetometry and neutron diffraction measurements, which are only sensitive to the bulk magnetic moment. As shown in Fig.~\ref{fig:ZFMuSR}(a-c) (see also \cite{SM}), we first carried out zero-field (ZF) measurements to investigate the magnetic ground state for $x$ = 0.05, 0.10, 0.25 and 0.30. Contrary to prior studies of Y$_{1-x}$Ca$_{x}$TiO$_{3}$ that reported the disappearance of magnetic order at $x = 0.20$ \cite{Tokura1992,Kumagai1993}, a clear precession/relaxation signal indicative of magnetism is observed at low temperature in $x = 0.25$ and 0.30. The ZF spectra are fit to the following function \cite{HameedYLa2021}:

	\begin{equation}
	\begin{split}
		A(t) = F\bigg(r \mathrm{e}^{-\lambda_{\text{T}}t}\cos{(2\pi \nu t + \phi)}
		\\+~(1-r) \mathrm{e}^{-\lambda_{\text{L}}t} \bigg)+~(1-F)\mathrm{e}^{-\lambda_{\text{p}}t}.
		\label{eq:1}
	\end{split}
	\end{equation}
	Here, $F$ is the magnetically ordered volume fraction, which is comprised of two components. The first term is an oscillating transverse component, associated with the component of the local field oriented perpendicular to the muon spin polarization and characterized by a precession frequency $\nu$, a phase offset $\phi$, and a damping rate $\lambda_{\text{T}}$ to allow for a distribution of the local fields. The second term is a slowly relaxing longitudinal component with a relaxation rate $\lambda_{\text{L}}$, that arises from the component of the local field oriented parallel to the muon spin polarization. The parameter $r$ ranges from 0 and 1, as the orientation of the local field with respect to the muon polarization direction changes from parallel to perpendicular \cite{HameedYLa2021}. Additionally, the non-oscillating component with relaxation rate $\lambda_{\text{p}}$ results from the PM volume fraction $(1-F)$. The total initial asymmetry and the asymmetry baseline were fixed to the values determined from a weak transverse field measurement in the high-temperature PM state of each sample. 
	The $x = 0.25$ and $0.30$ samples were measured with the DR spectrometer equipped with a dilution fridge, which uses a sample holder covered with silver. This adds an additional PM background signal from the parts of the holder that are not covered by the sample. Therefore, in these cases, an additional time-independent asymmetry arising from the silver was included in the fits. Furthermore, $r$ was set to 1, since the two samples were carefully oriented such that the $c$-axis along which the magnetic moment points is nearly perpendicular to the initial muon spin polarization direction. This helps to avoid the slowly relaxing longitudinal component from mixing with the background signal in the fits. Support for this assumption comes from the fact that the fitted value of $r$ is $\sim$ 0.9 for $x = 0.05$ and $x = 0.10$, where the $c$ axis was nearly perpendicular to the initial muon spin polarization direction. Whereas an oscillatory asymmetry component, indicative of a homogeneous static local field, is observed for $x = 0$ and 0.10, no oscillations are detected for $x = 0.25$ and 0.30. This indicates the absence of a homogeneous static local field at these Ca concentrations. However, a fast relaxation component is present for both $x = 0.25$ and 0.30, which could be driven by a distribution of static local fields, a time-dependent dynamic local field, or a combination of both \cite{Pula2024}. A good fit at low temperature for $x = 0.25$ and 0.30 could only be obtained by replacing the exponential in the transverse term with a Gaussian. This implies significant changes in the nature of the magnetism at these higher Ca concentrations, as discussed further below. Note that, at all Ca concentrations, the spectra above $T_C$ are well fit by a single exponentially-relaxing component, as expected for the PM phase.
	
		\begin{figure}
		\includegraphics[width=0.45\textwidth]{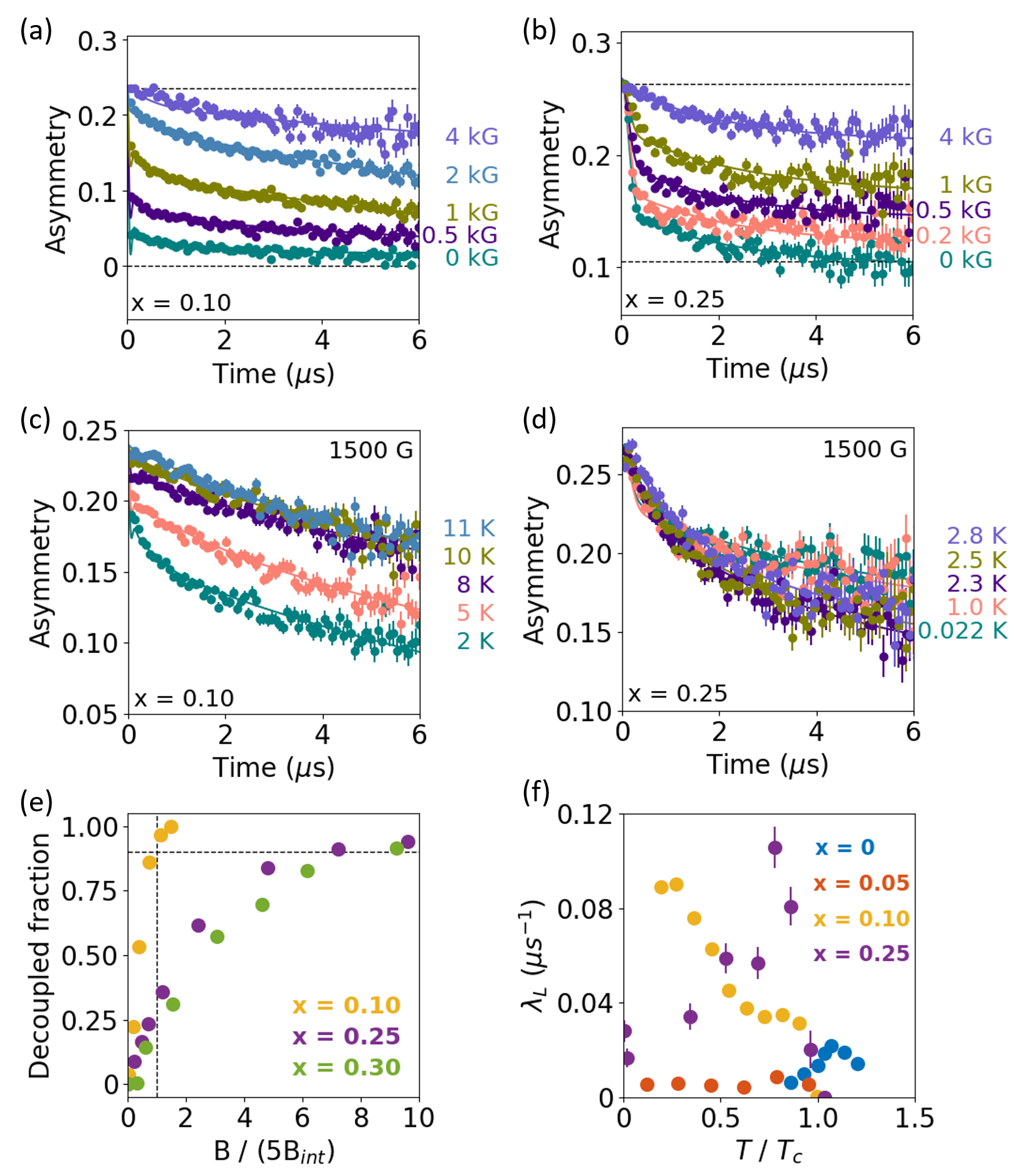}
		\caption{(a,b) LF $\mu$SR spectra at various fields for (a) $x = 0.10$ at 2 K and (b) $x = 0.25$ at 30 mK. The distance between the horizontal dashed lines represents the range that corresponds to the total asymmetry due to muons that stopped in the sample, excluding the background contribution. (c,d) $\mu$SR spectra at various temperatures in an applied LF of 1500 G for (c) $x = 0.10$, and (d) $x = 0.25$. Solid lines are fits to the data (see text). (e) The fraction of decoupled ZF relaxation $A_L$ as a function of the applied LF. Here, $B_{\text{int}}$ represents the internal field, which is determined from the ZF fits. The horizontal and vertical dashed lines correspond to 90\% decoupled fraction and applied LF $B = 5\times B_{\text{int}}$, respectively. (f) Temperature dependence of the dynamic relaxation rate $\lambda_L$. The data for $x = 0$ are reproduced from Ref.~\cite{HameedYLa2021}.}
		\label{fig:LFMuSR}
	\end{figure}

	Figure~\ref{fig:ZFMuSR}(d) displays the temperature and Ca concentration dependence of the static internal field, defined as $B_{\text{int}}$ = $\sqrt{(2\pi\nu)^2 + (\lambda_{T})^2}$ \cite{HameedYLa2021}. This encompasses both homogeneous and inhomogeneous contributions, which originate, respectively, from the well-defined average local field and the damping caused by the distribution of local fields around this average. The data for $x = 0$, reported previously in Ref.~\cite{HameedYLa2021}, are included for comparison. A strong suppression of the static local moment is observed with increasing Ca concentration.  Figure~\ref{fig:ZFMuSR}(e) displays the temperature and Ca concentration dependence of the magnetic volume fraction $F$. A significant reduction in the volume fraction, from nearly 100\% to approximately 50\%, is observed as the Ca content increases from $x = 0$ to $x = 0.30$. This provides clear evidence for a first-order phase transition. The observed reduction in both the local magnetic moment and the magnetic volume fraction with increasing Ca concentration suggests that the decrease in the bulk moment, as revealed by neutron diffraction and SQUID magnetometry (Fig.~\ref{fig:SQUID_neutron}(c)), arises from a combination of these effects. Figure~\ref{fig:ZFMuSR}(f) compares the Ca doping dependence of the bulk magnetic moment from $\mu$SR (calculated as the product of the local moment and the magnetic volume fraction) with the saturation magnetic moment derived from magnetometry. While both methods clearly demonstrate a suppression of bulk ordered moments with Ca doping, the bulk moment established from $\mu$SR exhibits a significantly stronger decrease. A similar observation was previously reported for Y$_{1-x}$La$_{x}$TiO$_{3}$ \cite{HameedYLa2021}, where it was attributed to the muon stoppage site being near the Ti$^{3+}$ ions \cite{Baker2008}. This proximity causes the $\mu$SR-measured local moment to be strongly influenced by subtle changes in the magnetic structure.
	
	\begin{figure}
		\includegraphics[width=0.48\textwidth]{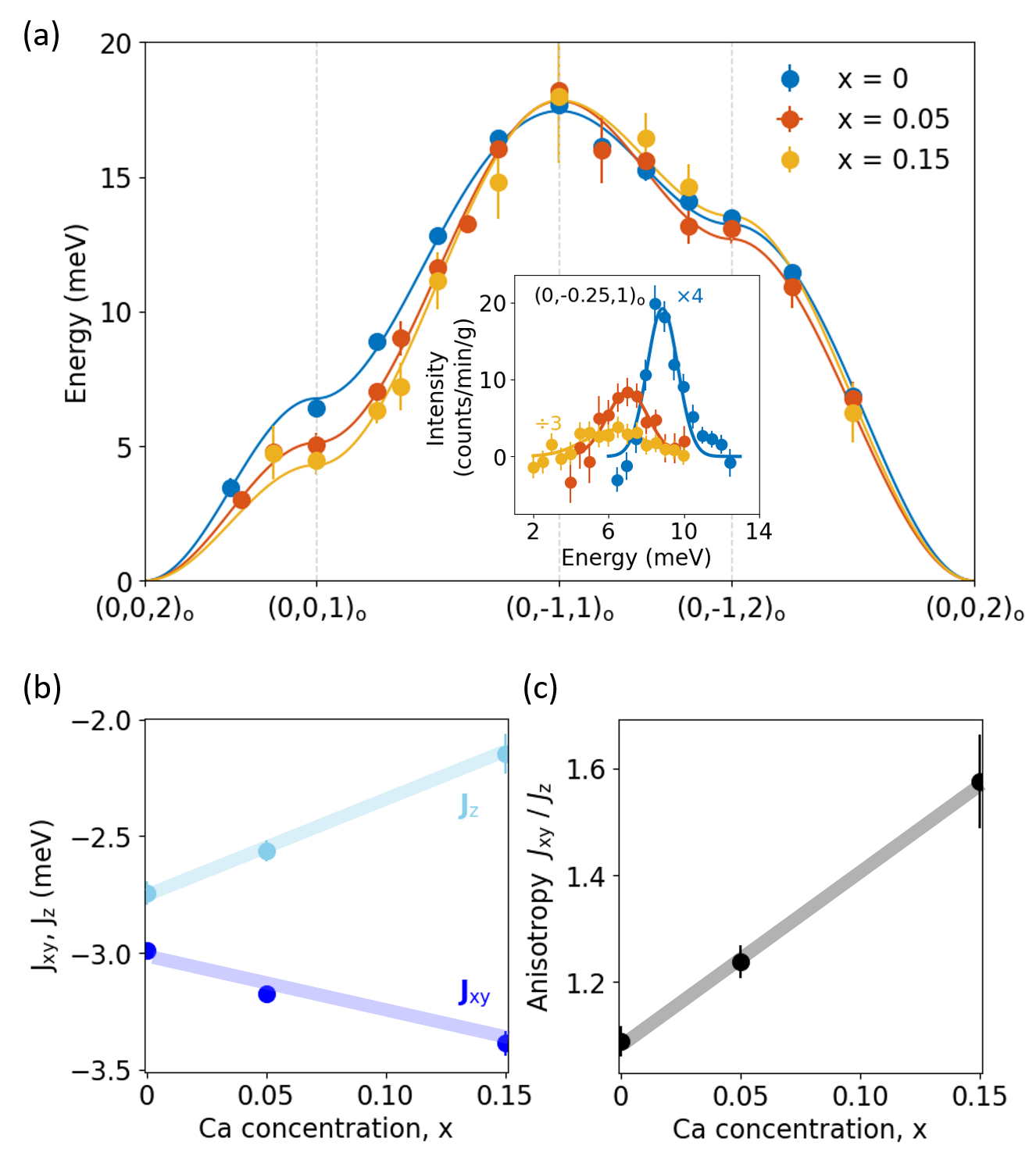}
		\caption{(a) Ca concentration dependence (up to $x = 0.15$) of the spin-wave spectra obtained at 2 K. The lines are fits to model spectra as described in the text. The data for undoped YTiO$_3$ are reproduced from Ref.~\cite{HameedYLaSW2023}.  The inset shows representative energy scans at (0,-0.25,1)$_\text{o}$ (with high-temperature data subtracted) for x = 0, 0.05 and 0.15, along with respective Gaussian fits. The data for $x = 0$ are scaled down by a factor of 4 and the data for $x = 0.15$ are scaled up by a factor of 3. (b) Spin-exchange constants $J_{xy}$ and $J_z$ obtained from the fits of the spectra in (a), and (c) their anisotropy $J_{xy}/J_z$. The subscript ``o'' in panel (a) refers to the orthorhombic notation. The lines is (b,c) are guides to the eye.}
		\label{fig:SW}
		
	\end{figure}

	In order to determine the static versus dynamic character of the precession signal observed in the ZF measurements, we performed measurements in a longitudinal field (LF) applied parallel to the intial muon spin polarization direction \cite{Goko2017,Pula2024}. If the precession originates from a homogeneous Gaussian distribution of static local fields, applying a LF five times stronger than the internal local field is expected to decouple the signal by approximately 90\% \cite{blundell2022muon}. As shown in Fig.~\ref{fig:LFMuSR}(a), such decoupling is clearly observed for $x = 0.10$ where the fast relaxation component at small timescales almost completely disappears at 2 kG which is $\sim$ $4 \times B_\text{int}$ for this sample. However, for $x = 0.25$ and $0.30$, the fast relaxation component remains substantial at $5 \times B_\text{int}$ $\sim$~300 - 400 G (Fig.~\ref{fig:LFMuSR}(b) and \cite{SM}). To quantify this behavior, we fit the LF data to:
	
	\begin{equation}
		\begin{split}
			A(t) = F\bigg(r(1-A_L) \mathrm{e}^{-\lambda_{\text{T}}t}\cos{(2\pi \nu t + \phi)}
			+~rA_L\mathrm{e}^{-\lambda_{\text{L}}t}
			\\+~(1-r) \mathrm{e}^{-\lambda_{\text{L}}t}
			 \bigg)+~(1-F)\mathrm{e}^{-\lambda_{\text{p}}t}.
			\label{eq:2}
		\end{split}
	\end{equation}
	Here, the parameters $F$, $r$, $\nu$, $\phi$, $\lambda_{\text{T}}$, $\lambda_{\text{L}}$ and $\lambda_{\text{p}}$ retain the same definitions as in Eq.~(\ref{eq:1}). The parameter $A_L$ quantifies the fraction of the fast relaxation component decoupled by the application of the LF. $F$, $r$, $\nu$, $\phi$, $\lambda_{\text{T}}$ and $\lambda_{\text{p}}$ for all LF data are fixed based on the fits of the ZF results, whereas $\lambda_{\text{L}}$ and $A_L$ are allowed to vary. Note that for $x = 0.30$, fitting the non-zero LF data required $\lambda_{\text{p}}$ to be fixed to a value less than that for the ZF data. This adjustment reflects a pronounced change in the slope of the PM long-time exponential (see \cite{SM}). The underlying reason for the effect of the LF on the PM relaxation rate, particularly at this Ca concentration, remains unclear. Figure~\ref{fig:LFMuSR}(e) shows the LF dependence of $A_L$ extracted from the fits. As noted, for $x = 0.10$, $A_L$ surpasses 0.9 at $\sim 5 \times B_\text{int}$. In contrast, for $x = 0.25$ and 0.30, $A_L$ only reaches $\sim 0.3$ at $5 \times B_\text{int}$. 
	
	In most $\mu$SR studies, the application of a LF provides an effective method to separate relaxation due to static random fields from dynamic processes due to spin fluctuations and/or spin waves.  When the relaxation of the transverse component $(1 - A_{L})$ of the asymmetry occurs significantly faster than the dynamic $1/T_{1}$ processes represented by the relaxation rate $\lambda_{L}$, it is safe to assume that the fast relaxation of the transverse component is predominantly due to static random fields. The time spectra in Fig.~\ref{fig:LFMuSR}(b), for example, are clearly consistent with this description.  In this case, the LF dependence of $A_{L}$ shown in Fig.~\ref{fig:LFMuSR}(e) should be attributed to the distribution of static random fields at the muon site. The decoupling as a function of LF  proceeds rapidly for a homogeneous local field, whereas it occurs more gradually in systems with a broader field distribution. This can be seen, for example, by comparing the LF dependence of $A_{L}$ in more homogeneous Gaussian \cite{Hayano1979} versus more random Lorentzian \cite{Uemura1985,Uemura1999} field distributions.  In the present case, involving phase separation for $x = 0.25$ and 0.30 (Fig.~\ref{fig:ZFMuSR}(e)), the field distribution can be even more random than Lorentzian. This would explain the slow dependence of $A_{L}$ on LF (Fig.~\ref{fig:LFMuSR}(e)) for $x = 0.25$ and 0.30.
	
	To investigate further, we carried out temperature-dependent LF measurements for several dopings (see Fig.~\ref{fig:LFMuSR}(c,d) and \cite{SM}) and fit the data to Eq.~\ref{eq:2}. Similar to the field dependence, only $\lambda_{L}$ and $A_L$ are allowed to vary, whereas the remaining parameters are fixed based of ZF data fits. The effect of dynamic processes can be seen from the $1/T_{1}$ relaxation rate $\lambda_{L}$ for the longitudinal component in Fig.~\ref{fig:LFMuSR}(f), represented by the second and third terms of Eq.~\ref{eq:2}.  In principle, the slow decay in LF can also include the effect of dynamic relaxation in the PM volume, given by the fourth term of Eq.~\ref{eq:2}. Significant dynamic relaxation can be seen for $x = 0.10$ and 0.25 at and below the transition temperature $T_C$. Below, we provide a detailed comparison of the present results with $\mu$SR studies on various Mott transition systems and weak FM systems with itinerant electrons.

	\subsection{Inelastic neutron scattering}
	\label{Sec:INS}
	Recent ESR measurements of YTiO$_3$ revealed a low-lying $\sim$ 10 meV orbital gap that decreases nearly linearly with the decrease of $T_C$ induced by Ca doping \cite{Najev2024}. Since changes in the orbital state are expected to be reflected in the spin-exchange Hamiltonian, we measured spin waves via inelastic neutron scattering. Figure~\ref{fig:SW} displays the spin-wave spectra for $x = 0.05$ and 0.15, along with prior data for $x = 0$ \cite{HameedYLa2021} for comparison. Inelastic neutron scattering measurements were not performed for $x > 0.15$, due to the much weaker spin-wave intensities (see inset of Fig.~\ref{fig:SW}(a) and \cite{SM}). Overall, no significant changes were observed in the zone-boundary spin-wave energy, although $T_C$ for $x = 0.15$ \cite{Najev2022} is about a factor of seven lower than that for $x = 0$. DFT calculations have shown that La substitution can lead to an easy-plane spin-exchange anisotropy \cite{HameedYLaSW2023}. Since Ca doping increases the average Ti-O-Ti bond angle \cite{Tsubota2004}, similar to isovalent La substitution, we fit the spin-wave spectra to the following Hamiltonian: 
	
\begin{equation}
	\hat{H} = \sum_{<i,j>_{a,b}}J_{xy}\bar{S_i} \cdot \bar{S_j} + \sum_{<i,j>_{c}}J_{z}\bar{S_i} \cdot \bar{S_j}.
	\label{eq:3}
\end{equation}
Using linear spin-wave theory and the Holstein-Primakoff transformations, the spin-wave dispersion is $\hbar\omega = -2J_{z}S(1-\gamma_{z}) - 4J_{xy}S(1-\gamma_{xy})$, with $\gamma_{z} = \cos q_{z}c', \gamma_{xy} = \frac{1}{2}(\cos q_{x}a' + \cos q_{y}b')$.
Here, $a',b'$, and $c'$ are the pseudocubic lattice parameters, related to the orthorhombic lattice parameters via $a' = a/\sqrt{2}, b' = b/\sqrt{2},$ and $c' = c/2$.
The spin-exchange parameters obtained from the fits are shown in Fig.~\ref{fig:SW} (b, c). A pronounced easy-plane anisotropy emerges with increasing Ca content, indicating significant changes in the orbital state upon doping YTiO$_3$ with Ca. 

\section{Discussion}

\begin{table*}[t]
	\centering
	\renewcommand{\arraystretch}{1.5}  
	\LTcapwidth=\textwidth  
	\caption{Comparison of the present results for (Y,Ca)TiO$_3$ with $\mu$SR results for various Mott and itinerant weak FM systems.}
	\label{table1}
	\begin{tabular}{|>{\centering\arraybackslash}m{1.8cm}|>{\centering\arraybackslash}m{2.5cm}|>{\centering\arraybackslash}m{2.5cm}|>{\centering\arraybackslash}m{2.5cm}|>{\centering\arraybackslash}m{2.2cm}|>{\centering\arraybackslash}m{2.5cm}|>{\centering\arraybackslash}m{1.5cm}|}
		\hline
		\textbf{Material} & \textbf{Tuning} & \textbf{Phase transitions} & \textbf{Structural transition} & \textbf{Magnetic phase separation} & \textbf{Spin fluctuations ($\mu$SR timescale)} & \textbf{Reference} \\ 
		\hline
		\multicolumn{7}{|c|}{\textbf{Mott systems}} \\ \hline
		RENiO$_3$ & RE ion (tilt angle) & AFI-PMM & Yes & Yes, strong & Minimal & \cite{Frandsen2016} \\ \hline
		V$_2$O$_3$ & Pressure & AFI-PMM & Yes & Yes, strong& Minimal & \cite{Frandsen2016} \\ \hline
		Ni(S,Se)$_2$ & (S,Se) isovalent substitution &  AFI-AFM-PMM &  No &  Yes, weak & Minimal & \cite{Sheng2022} \\ \hline
		BaCoS$_2$ & Pressure &  AFI-AFM-PMM &  No &  Yes, strong & Minimal & \cite{Guguchia2019} \\ \hline
		Ba(Co,Ni)S$_2$ & (Co,Ni) charge doping & AFI-AFM-PMM &  No &  Yes, strong & Minimal & \cite{Guguchia2019} \\ \hline
		(Y,La)TiO$_3$ & (Y,La) isovalent substitution &  FMI-AFI  & No &  Yes, weak & Minimal & \cite{HameedYLa2021} \\ \hline
		(Y,Ca)TiO$_3$ & (Y,Ca) charge doping & FMI-PMM & Yes \cite{Kato2002} &  Yes, strong & Modest & Present work \\ \hline
		\multicolumn{7}{|c|}{\textbf{Itinerant weak FM systems}} \\ \hline
		MnSi & Pressure &Helimagnetic metal-PMM & No &  Yes, strong & Minimal & \cite{Uemura2007} \\ \hline
		(Mn,Fe)Si & Pressure & Helimagnetic metal-PMM &  No & No, second-order & Critical & \cite{Goko2017} \\ \hline
	\end{tabular}
\end{table*}

\begin{figure}
	\includegraphics[width=0.48\textwidth]{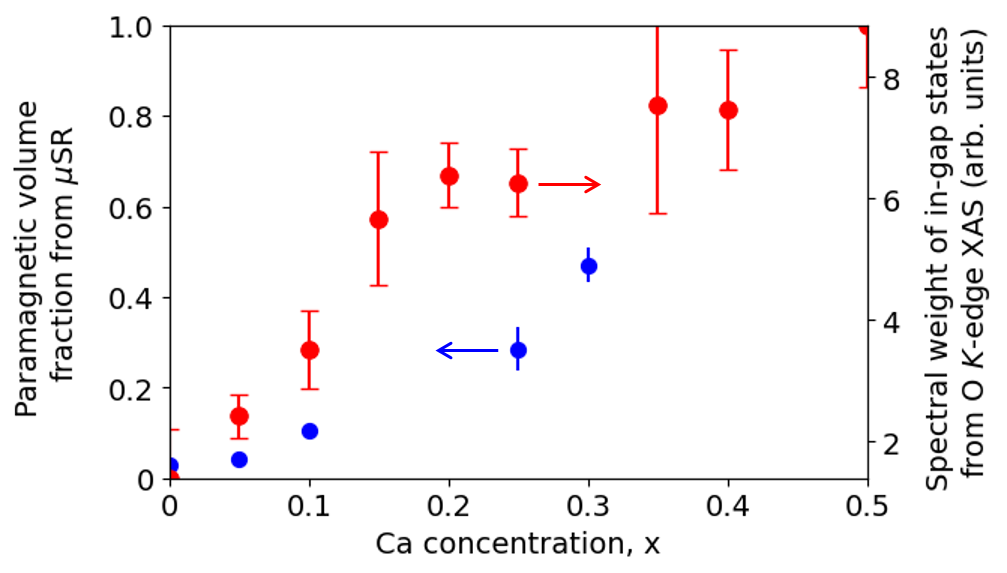}
	\caption{Ca-concentration dependence of the PM volume fraction measured by $\mu$SR below $T_C$, compared with the spectral weight of in-gap states estimated from O $K$-edge XAS (peak A in Ref.~\cite{HameedYCa2021}). The latter serves as a measure of the volume fraction of the hole-rich metallic phase \cite{HameedYCa2021}.}
	\label{fig:VF}
\end{figure}

Our $\mu$SR results clearly show that the FM-PM transition in Y$_{1-x}$Ca$_{x}$TiO$_3$ occurs via volume-wise phase separation into FM and PM regions, with the PM volume fraction reaching approximately 50\% for $x = 0.30$ (see Fig.~\ref{fig:VF}). Recent work has shown that the Ca-doped system electronically phase-separates into hole-poor Mott insulating and hole-rich metallic regions \cite{HameedYCa2021}, which naturally leads to the coexistence of magnetically ordered and PM volumes. X-ray absorption spectra (XAS) at the Ti $L$-edge and O $K$-edge were shown to be consistent with an increasing (decreasing) volume fraction of the hole-rich metallic (hole-poor insulating) phase. Figure~\ref{fig:VF} shows the Ca-concentration dependence of the PM volume fraction inferred from $\mu$SR, as well as the spectral weight of in-gap states (peak A in Ref.\cite{HameedYCa2021}) obtained from O $K$-edge XAS. Since this spectral weight serves as a measure of the metallic volume fraction \cite{HameedYCa2021}, the close tracking of its $x$-dependence with that of the PM volume fraction indicates that the PM phase corresponds to the hole-rich metallic regions. Accurate quantification of the metallic and insulating volume fractions with XAS is challenging, which likely contributes to the observed discrepancies in Fig.~\ref{fig:VF}. Theoretical models that correctly predict the occurrence of electronic phase separation in RETiO$_3$ also predict the emergence of an intermediate magnetic metallic phase prior to the transition to a PM metal \cite{Yee2015,Kurdestany2017}. Given that the FM order persists up to $x = 0.30$, it remains a strong possibility that such a magnetic metallic phase exists within the doping range $x = 0.30$ to 0.35.

In previous $\mu$SR studies of systems near the disappearance of static magnetism, a few different types of responses have been observed: (a) emergence of phase separation and suppression of dynamic effects at the quantum phase transition in MnSi under pressure and in (Sr,Ca)RuO$_3$ \cite{Uemura2007}; (b) disappearance of phase separation and recovery of second-order transition due to disorder in (Mn,Fe)Si under pressure \cite{Goko2017}; and (c) spin dynamics persistent at $T \rightarrow 0$ with an anomalous ``hardly decouplable'' line shape in spin liquid candidate SrCr$_8$Ga$_4$O$_{19}$ \cite{Uemura1994}. The present case of (Y,Ca)TiO$_3$ exhibits significant dynamics with phase separation, but without a signature of dynamics persisting to $T \rightarrow 0$.  This is a new type of response that differs from the above mentioned cases (a)-(c). Further studies will be needed to clarify the essence of the present behavior in (Y,Ca)TiO$_3$, which is related to the underlying insulator-metal transition and phase separation associated with inhomogeneity in charge distributions \cite{HameedYCa2021}. 

\begin{figure}
	\includegraphics[width=0.48\textwidth]{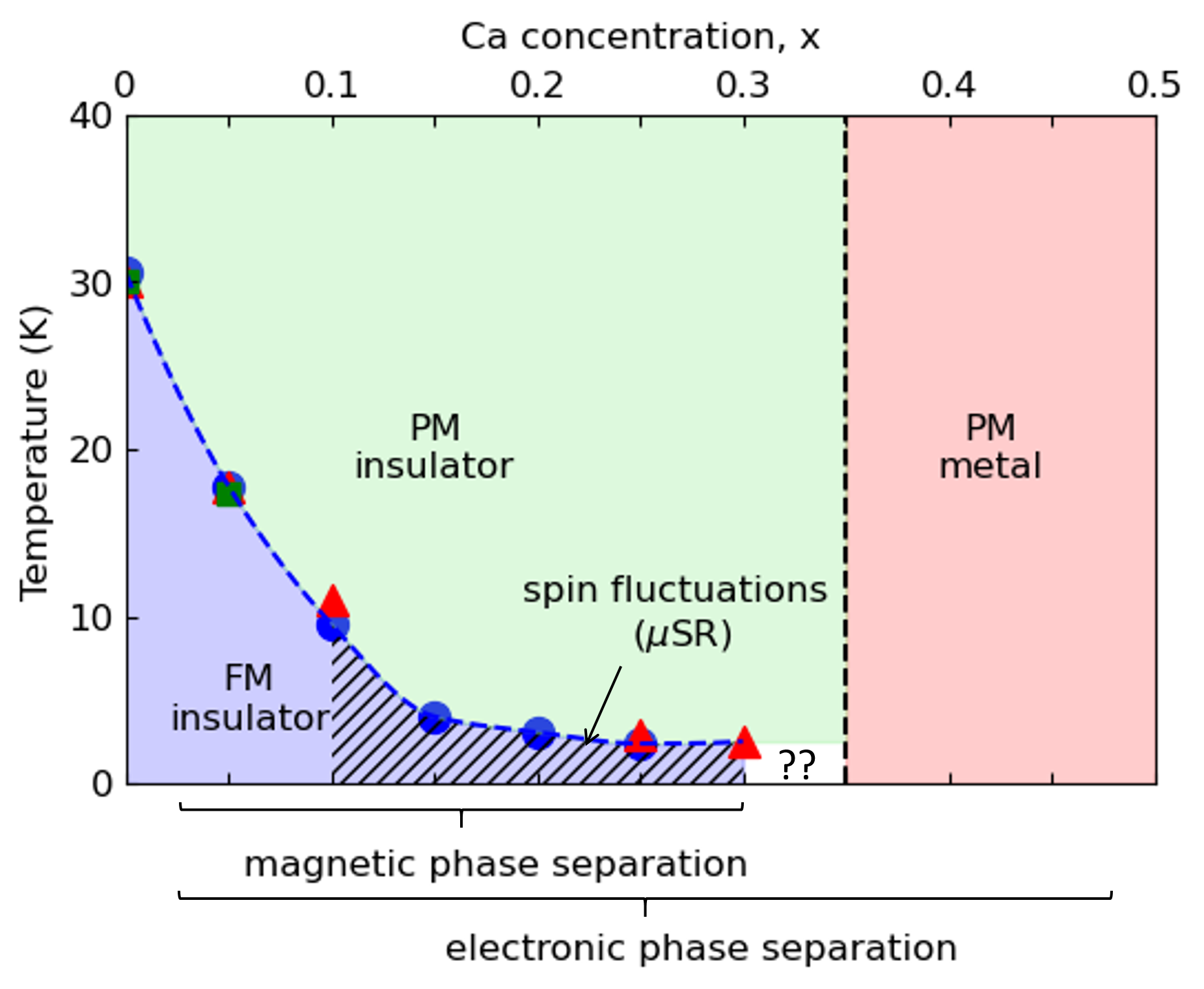}
	\caption{Phase diagram of Y$_{1-x}$Ca$_{x}$TiO$_3$. The $T_C$ values extracted from the AC susceptibility data in Fig.~\ref{fig:SQUID_neutron}(a) are represented by blue circles, those obtained from fits to the neutron diffraction data in Fig.~\ref{fig:SQUID_neutron}(d) are shown as green squares, and those corresponding to the transition midpoints in the $\mu$SR data from Fig.~\ref{fig:ZFMuSR}(e) are indicated by red triangles. Spin fluctuations at the $\mu$SR timescale are observed at and below $T_C$ for $x \geq 0.10$. Electronic \cite{HameedYCa2021} and magnetic phase separation are observed over a wide range of Ca concentrations.}
	\label{fig:PD}
\end{figure}

$\mu$SR studies of static and dynamic magnetism near the disappearance of static magnetic order have been performed in several other Mott transition systems and in a few weak ferro/helimagnets with itinerant electrons \cite{HameedYLa2021,Goko2017,Frandsen2016,Sheng2022,Guguchia2019,Uemura2007}. Table~\ref{table1} compares these results with the present findings for (Y,Ca)TiO$_3$. The existence of phase separation with a suppressed dynamic response has been observed in all Mott transition systems, including the single-step transition from an AF insulator (AFI) to a paramagnetic metal (PMM) in RENiO$_3$ \cite{Frandsen2016} and V$_2$O$_3$ \cite{Frandsen2016}, as well as the two-step Mott transition from an AFI to an AF metal (AFM) and then to a PMM in Ni(S,Se)$_2$ \cite{Sheng2022} and Ba(Co,Ni)S$_2$ \cite{Guguchia2019}. The present results for (Y,Ca)TiO$_3$ with phase separation and without critical spin dynamics are generally consistent with the behaviors found in these other Mott transition systems. The existence of a modest dynamic response in the present system at high Ca concentrations might be related to built-in spatial randomness, which was found to cause the recovery of a second-order transition and critical dynamics in the evolution from disorder-free MnSi \cite{Uemura2007} to disorder-rich (Mn,Fe)Si \cite{Goko2017}. The relationship with possible randomness in the charge conduction channel, which could be studied by scanning tunneling microscopy or other spatially resolving probes, would be an interesting future research direction. However, the surface oxidation effect associated with the instability of the Ti$^{3+}$ state may pose significant challenges for surface-sensitive studies (see \cite{Aeschlimann2018,Aeschlimann2024} and SM of Ref.~\cite{HameedYCa2021}).

With regard to the observed easy-plane anisotropy in the spin-exchange parameters with increasing Ca doping in Section~\ref{Sec:INS}, it is important to note that orbital fluctuations typically involving equal contributions from all three $t_{2g}$ orbitals are expected to produce isotropic spin-exchange interactions \cite{Khaliullin2002, Khaliullin2003, Khaliullin2005}. In YTiO$_3$, the $d$-type Jahn-Teller distortion lifts the three-fold degeneracy of the $t_{2g}$ orbitals, resulting in a lower-energy two-fold degenerate set of orbitals and a single higher-lying orbital. The orbital gap detected by ESR corresponds to the splitting between the two low-lying orbitals \cite{Najev2024}. Consequently, the orbital fluctuations associated with the decreasing gap in doped Y$_{1-x}$Ca$_{x}$TiO$_{3}$ are expected to involve contributions from only two $t_{2g}$ orbitals, consistent with the anisotropic spin-exchange parameters identified in our study. Nevertheless, it remains a theoretical challenge to explain why undoped YTiO$_3$ exhibits nearly isotropic spin-exchange interactions despite the complete lifting of the orbital degeneracy (see Fig.~\ref{fig:SW}(b,c) and Ref.~\cite{Ulrich2002}). Notably, prior first-principles calculations predict an easy-plane anisotropy in spin-exchange for undoped YTiO$_3$, with the calculated anisotropy $J_\text{xy}$/$J_z$ ranging from 1.5 to 8 \cite{Solovyev2004, Pavarini2005, HameedYLaSW2023}.

\section{Conclusions}
	
Figure~\ref{fig:PD} summarizes our main findings in the form of a revised phase diagram for Y$_{1-x}$Ca$_{x}$TiO$_3$. Ferromagnetism persists up to a Ca concentration of at least $x = 0.30$, and the FM-PM transition occurs via a volume-wise phase separation into FM ordered and PM regions. The $x$-dependence of the PM volume fraction closely follows that of the volume fraction of the hole-rich metallic phase. Spin fluctuations with a characteristic timescale of $\sim$ 0.1 $\mu$s appear at and below $T_C$ for Ca doping levels $x \geq 0.10$, as detected via $\mu$SR. The observed magnetic phase separation, accompanied by a modest dynamic response, reveals a distinct behavior among Mott systems approaching the loss of static magnetism, likely driven by the underlying insulator-metal transition and the resulting electronic phase separation. Finally, our spin-wave measurements reveal a pronounced anisotropy in spin-exchange parameters in the doped system, indicative of substantial changes in the orbital state compared to undoped YTiO$_3$.
	
\section{Acknowledgments}
The work at University of Minnesota was funded by the Department of Energy through the University of Minnesota Center for Quantum Materials, under DE-SC0016371. The work at Columbia University was supported by US NSF grant DMR 2104661. Parts of this work were carried out in the Characterization Facility, University of Minnesota, which receives partial support from the NSF through the MRSEC (Award Number DMR-2011401) and the NNCI (Award Number ECCS-2025124) programs. A portion of this research used resources at the High Flux Isotope Reactor, a DOE Office of Science User Facility operated by the Oak Ridge National Laboratory.

\bibliographystyle{apsrev4-2}

\bibliography{YCTO.bib}

\widetext
\clearpage

\begin{center}
	\textbf{\large Supplemental Material}
\end{center}

Here we document additional $\mu$SR data and raw spin-wave data at all measured Ca concentrations.

\setcounter{equation}{0}
\setcounter{section}{0}
\setcounter{figure}{0}
\setcounter{table}{0}
\setcounter{page}{1}
\makeatletter
\renewcommand{\theequation}{S\arabic{equation}}
\renewcommand{\thetable}{S\arabic{equation}}
\renewcommand{\thefigure}{S\arabic{figure}}
\renewcommand{\thesection}{S\arabic{section}}
\renewcommand{\bibnumfmt}[1]{[S#1]}

\section{Additional $\mu$SR data}
\label{rawspinwave}

In this Section, we present additional $\mu$SR data that were not presented in the main text.

\begin{figure}[!htb]
	\includegraphics[width=0.9\textwidth]{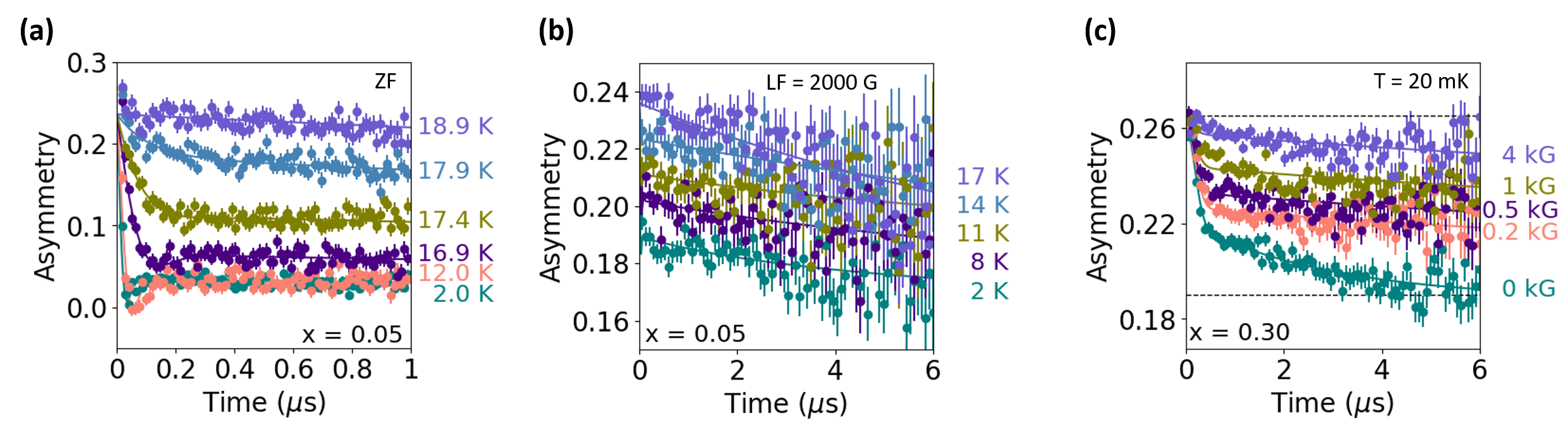}
	\caption{(a) ZF $\mu$SR spectra at different temperatures for $x = 0.05$. (b) $\mu$SR spectra at different temperatures in an applied LF of 2000 G for $x = 0.05$. (c) $\mu$SR spectra obtained with different applied LF for $x = 0.30$ at 20 mK. The distance between the horizontal dashed lines in (c) represents the range that corresponds to the total asymmetry due to muons that stopped in the sample, excluding the background contribution. Solid lines in (a-c) are fits to the data as described in the main text.}
	\label{fig:YCa05_00L_3meV}
\end{figure}

\section{Raw spin-wave data}
\label{rawspinwave}

In this Section, we present raw energy and momentum scans used to determine spin-wave dispersions presented in the main text. In all cases, the difference between a low-temperature scan and a high-temperature scan is fit to a simple gaussian profile. For each case, two separate plots are presented, one with the raw low-temperature (blue filled circles) and high-temperature (red filled circles) scans, and one with the difference data (black filled circles) fitted to a gaussian profile (solid black line). In all figures in this Section, the errorbars correspond to one standard deviation.

\subsection{$x = 0.05$}

\begin{figure}[!htb]
	\includegraphics[width=0.7\textwidth]{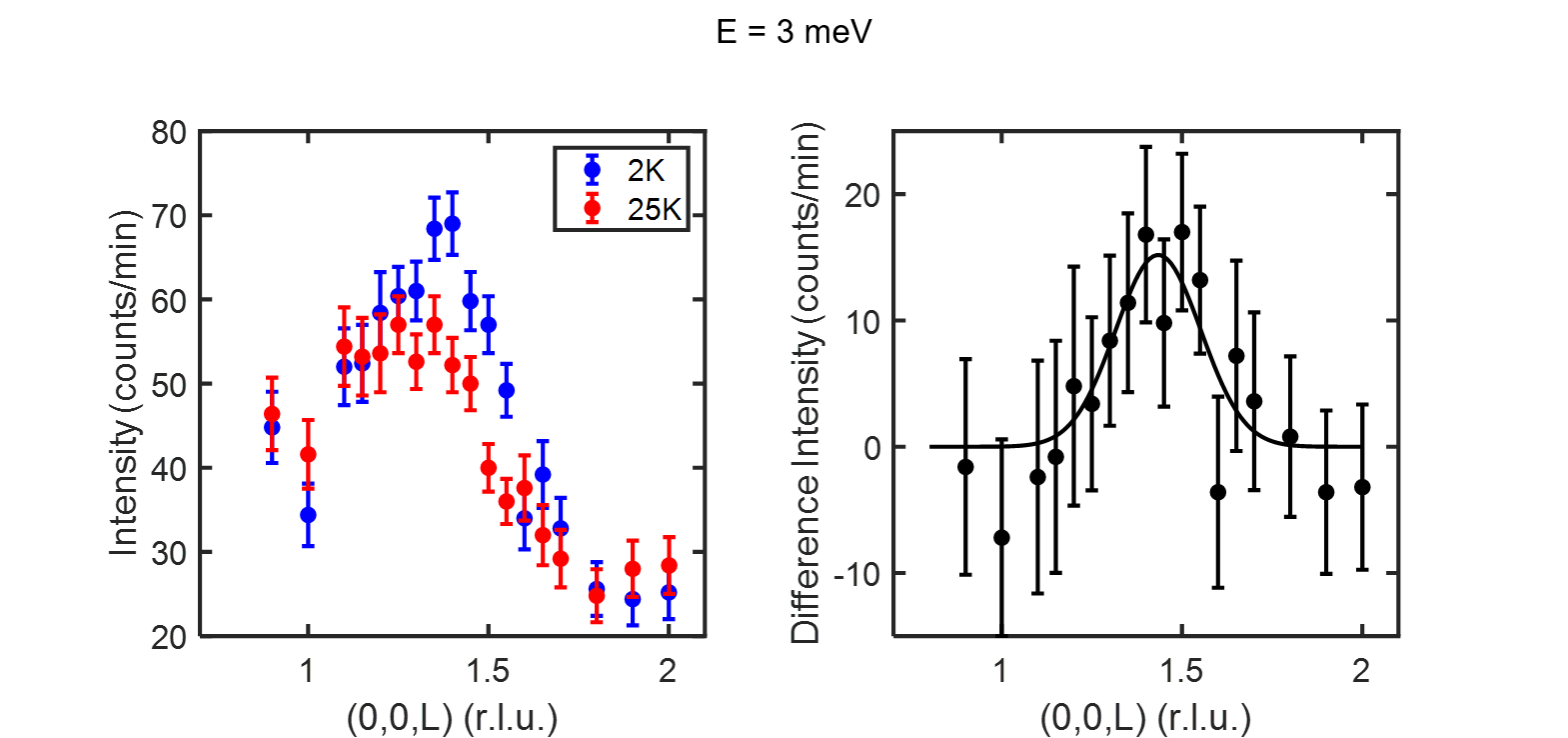}
	\caption{Left: Raw energy scans for $x = 0$ and $(0,0,1.5)_\text{o}$ at 1.5 K (red circles) and 40 K (blue circles). Right: Difference between the raw scans (black circles) and fit to gaussian profile (solid black line).}
	\label{fig:YCa05_00L_3meV}
\end{figure}

\begin{figure}[!htb]
	\includegraphics[width=0.7\textwidth]{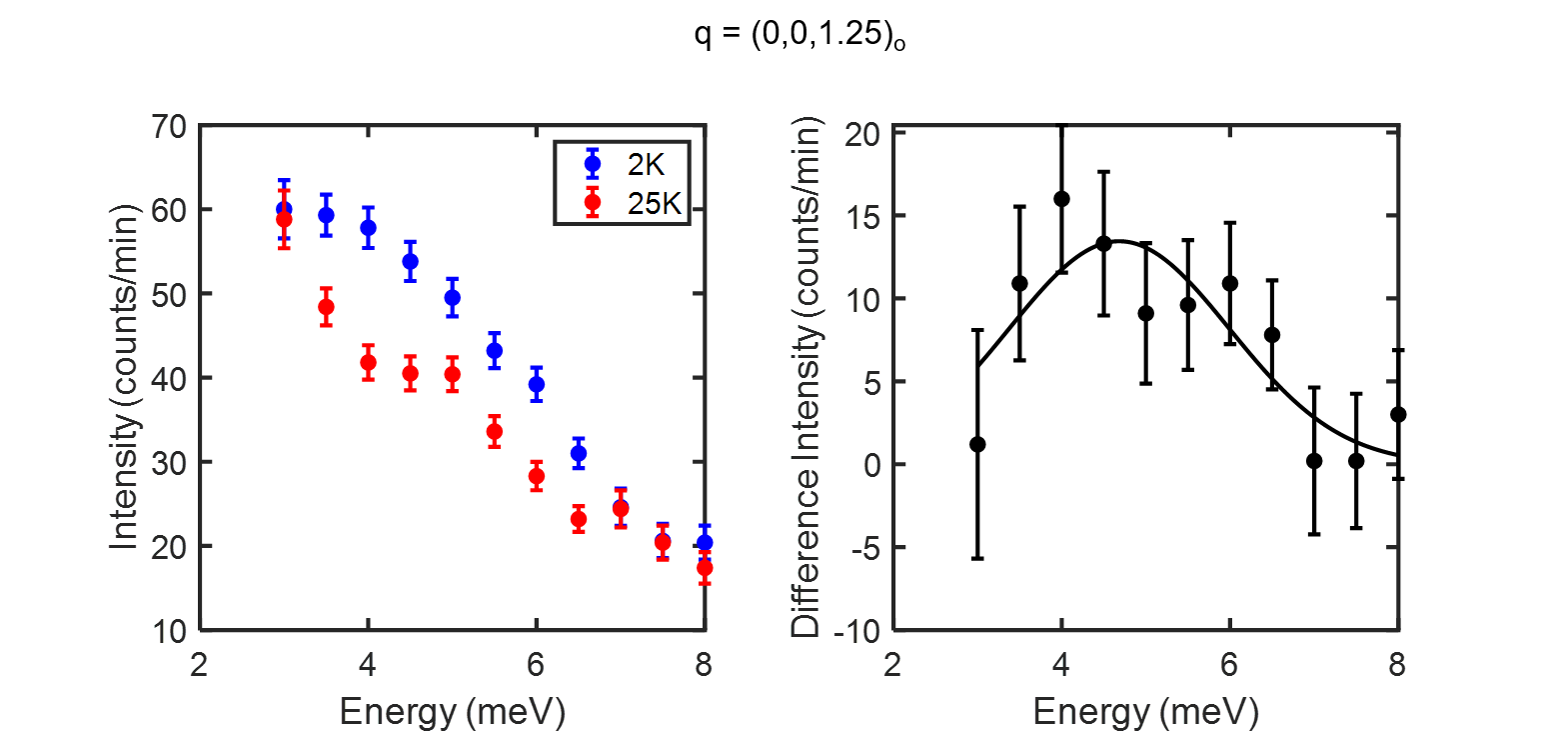}
	\caption{Left: Raw energy scans for $x = 0.05$ and $(0,0,1.25)_\text{o}$ at 2 K (blue circles) and 25 K (red circles). Right: Difference between the raw scans (black circles) and fit to gaussian profile (solid black line).}
	\label{fig:YCa05_001P25}
\end{figure}

\begin{figure}[!htb]
	\includegraphics[width=0.7\textwidth]{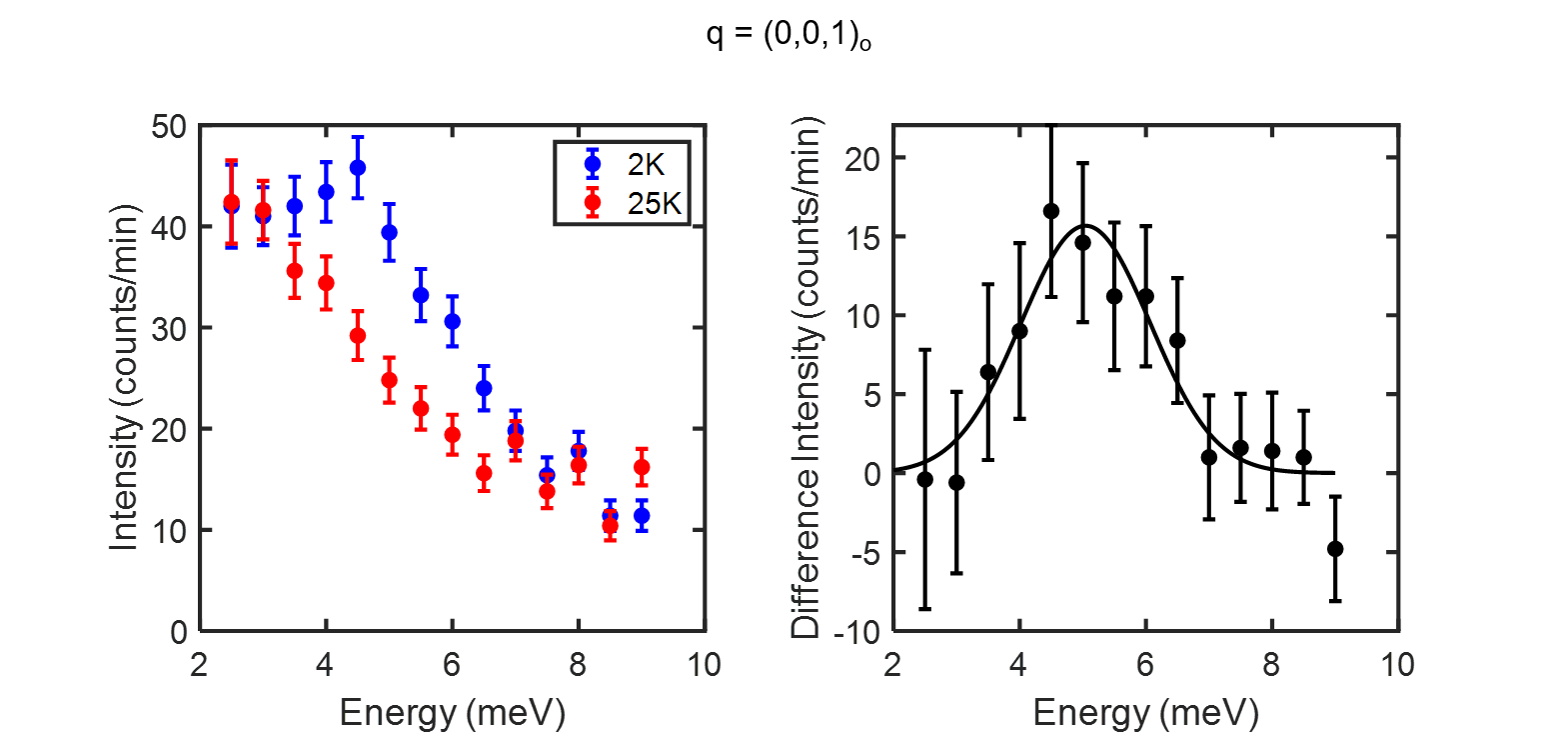}
	\caption{Left: Raw energy scans for $x = 0.05$ and $(0,0,1)_\text{o}$ at 2 K (blue circles) and 25 K (red circles). Right: Difference between the raw scans (black circles) and fit to gaussian profile (solid black line).}
	\label{fig:YCa05_001}
\end{figure}

\begin{figure}[!htb]
	\includegraphics[width=0.7\textwidth]{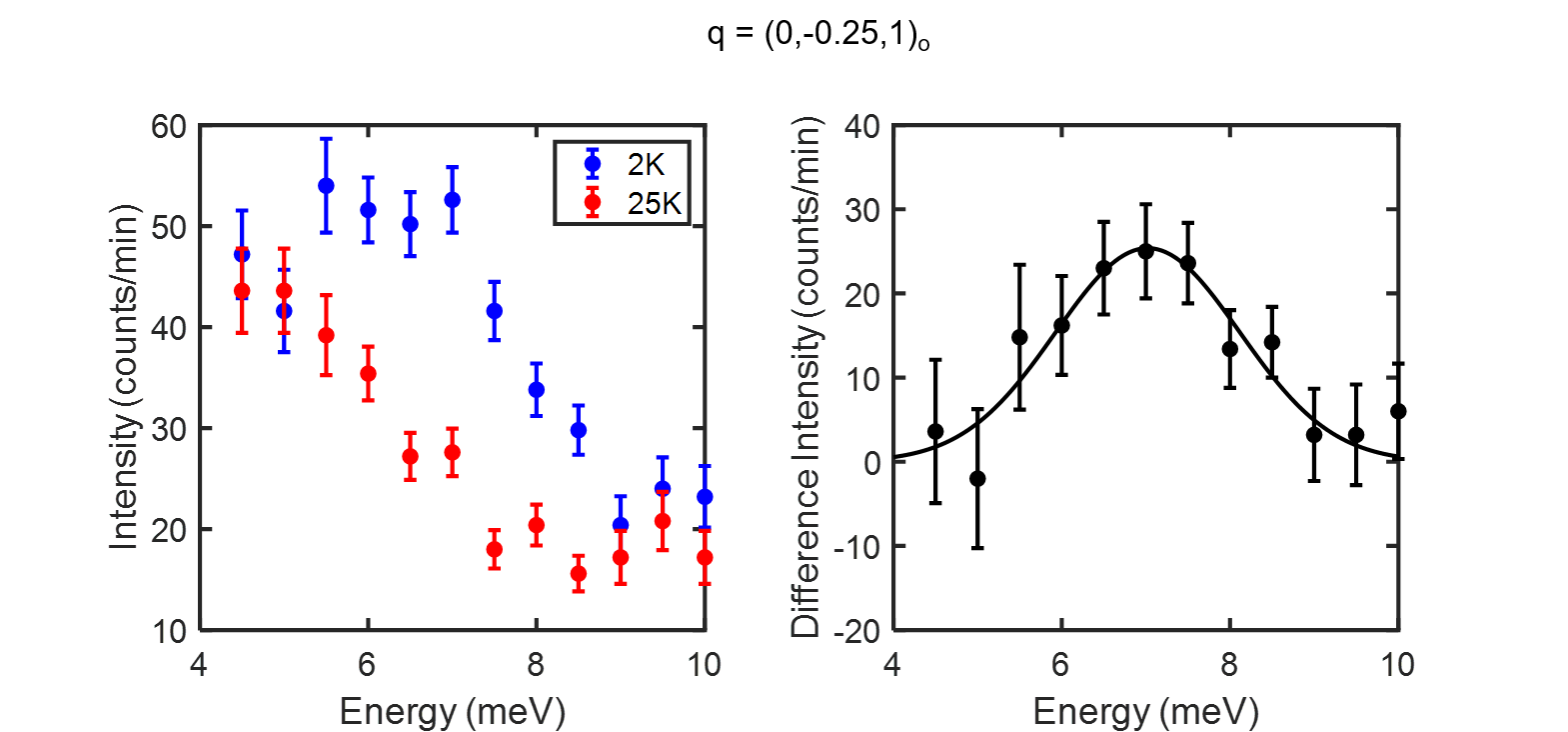}
	\caption{Left: Raw energy scans for $x = 0.05$ and $(0,-0.25,1)_\text{o}$ at 2 K (blue circles) and 25 K (red circles). Right: Difference between the raw scans (black circles) and fit to gaussian profile (solid black line).}
\end{figure}

\begin{figure}[!htb]
	\includegraphics[width=0.7\textwidth]{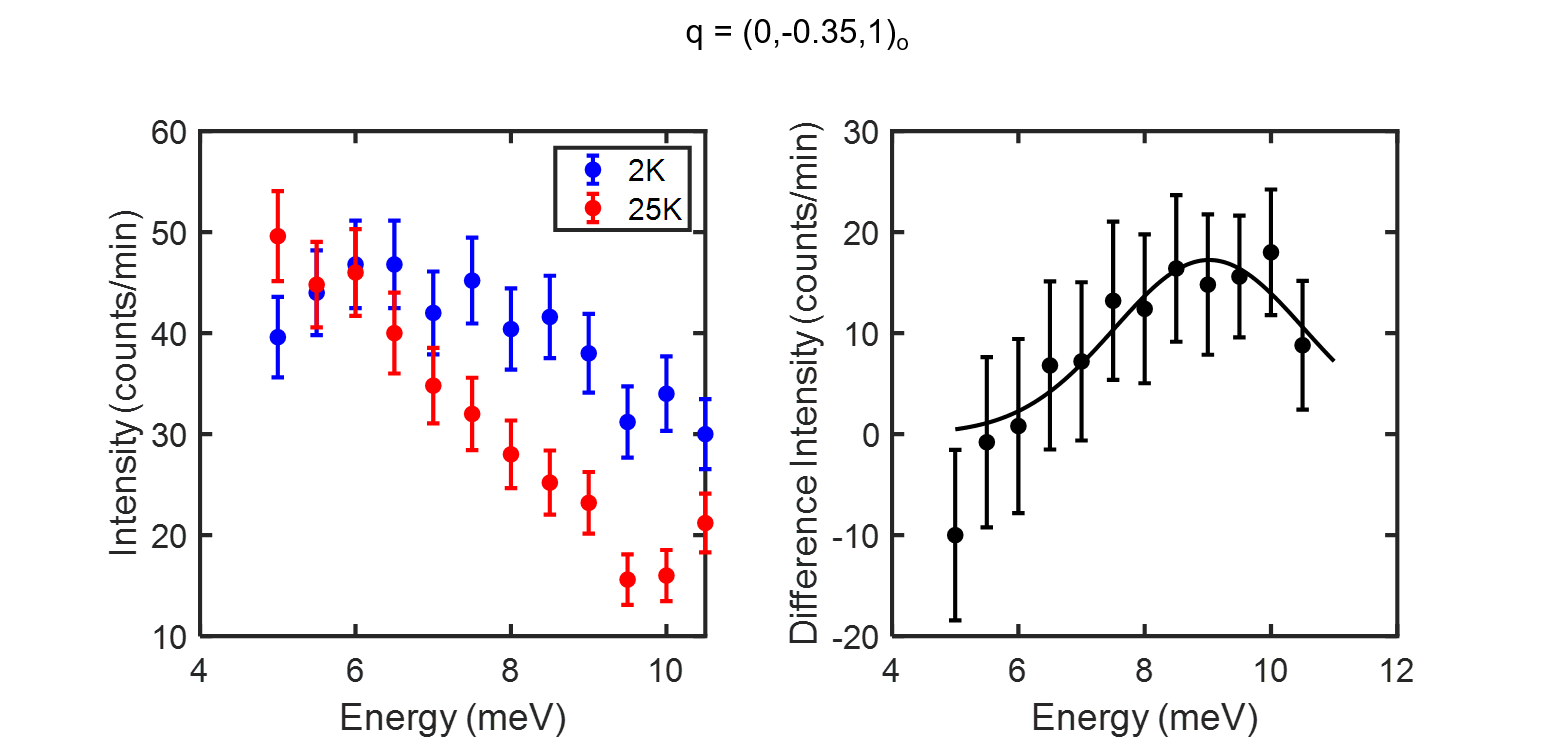}
	\caption{Left: Raw energy scans for $x = 0.05$ and $(0,-0.35,1)_\text{o}$ at 2 K (blue circles) and 25 K (red circles). Right: Difference between the raw scans (black circles) and fit to gaussian profile (solid black line).}
	\label{fig:YCa05_00P351}
\end{figure}

\begin{figure}[!htb]
	\includegraphics[width=0.7\textwidth]{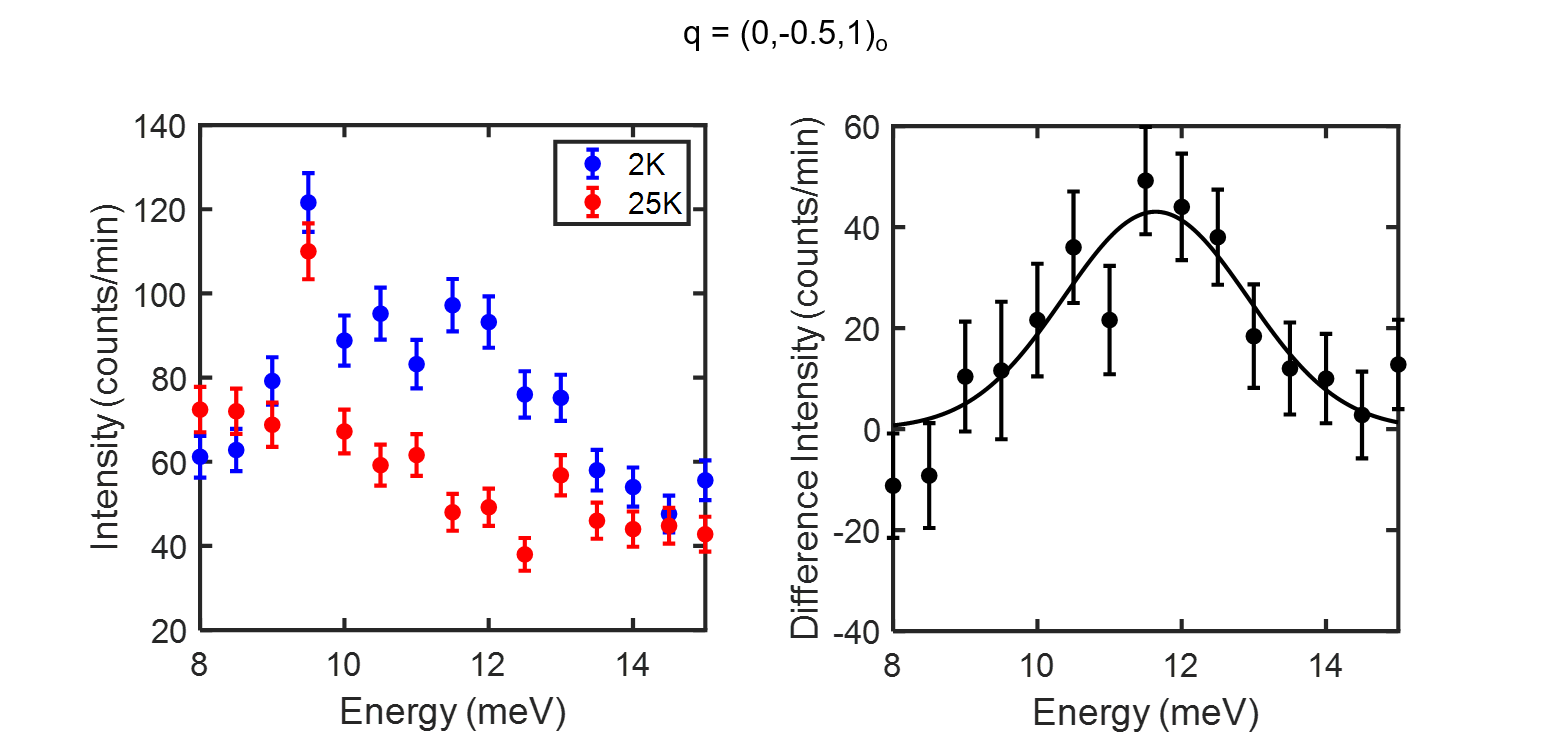}
	\caption{Left: Raw energy scans for $x = 0.05$ and $(0,-0.5,1)_\text{o}$ at 2 K (blue circles) and 25 K (red circles). Right: Difference between the raw scans (black circles) and fit to gaussian profile (solid black line).}
	\label{fig:YCa05_00P51}
\end{figure}

\begin{figure}[!htb]
	\includegraphics[width=0.7\textwidth]{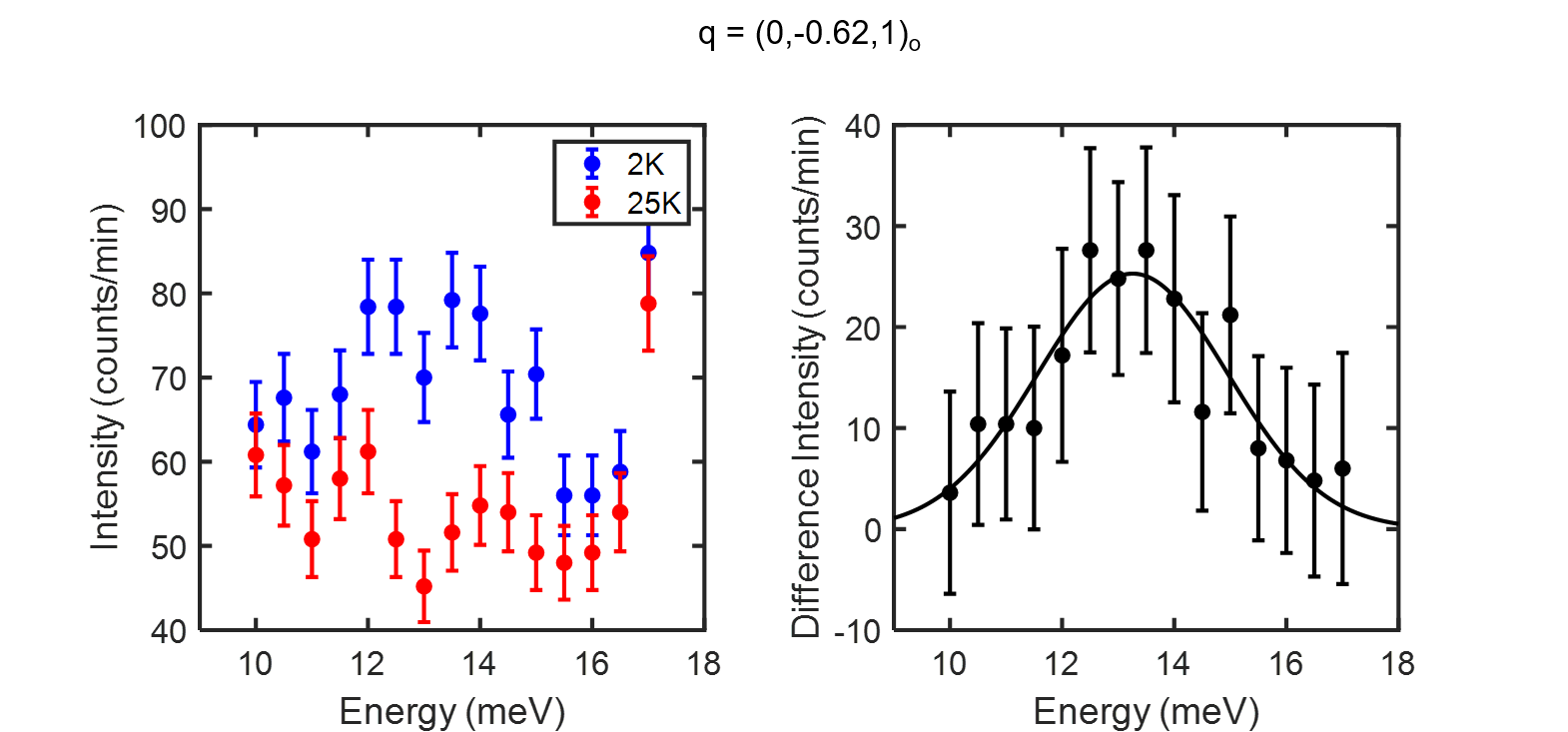}
	\caption{Left: Raw energy scans for $x = 0.05$ and $(0,-0.62,1)_\text{o}$ at 2 K (blue circles) and 25 K (red circles). Right: Difference between the raw scans (black circles) and fit to gaussian profile (solid black line).}
	\label{fig:YCa05_00P621}
\end{figure}

\begin{figure}[!htb]
	\includegraphics[width=0.7\textwidth]{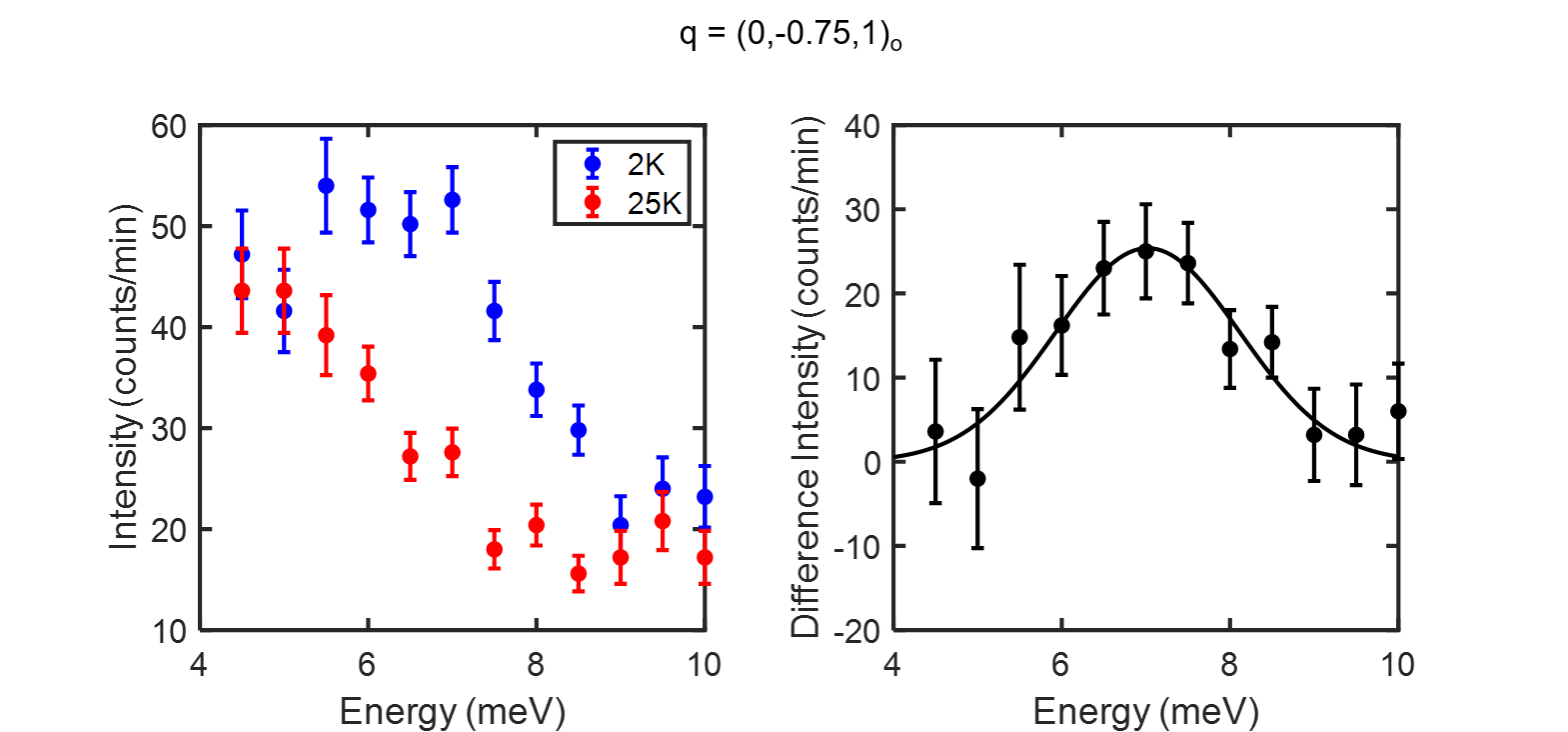}
	\caption{Left: Raw energy scans for $x = 0.05$ and $(0,-0.75,1)_\text{o}$ at 2 K (blue circles) and 25 K (red circles). Right: Difference between the raw scans (black circles) and fit to gaussian profile (solid black line).}
	\label{fig:YCa05_00P751}
\end{figure}

\begin{figure}[!htb]
	\includegraphics[width=0.7\textwidth]{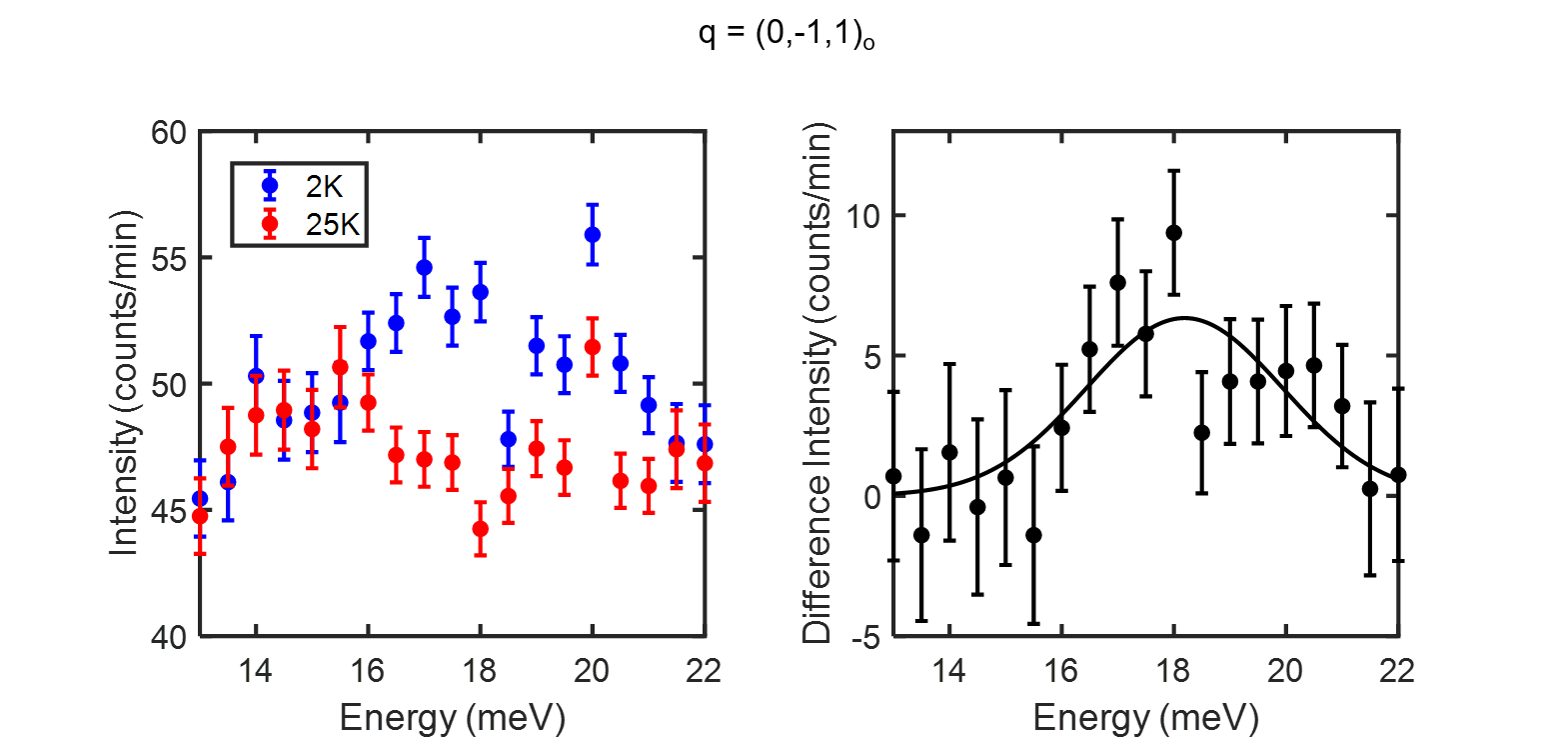}
	\caption{Left: Raw energy scans for $x = 0.05$ and $(0,-1,1)_\text{o}$ at 2 K (blue circles) and 25 K (red circles). Right: Difference between the raw scans (black circles) and fit to gaussian profile (solid black line).}
	\label{fig:YCa05_011}
\end{figure}

\begin{figure}[!htb]
	\includegraphics[width=0.7\textwidth]{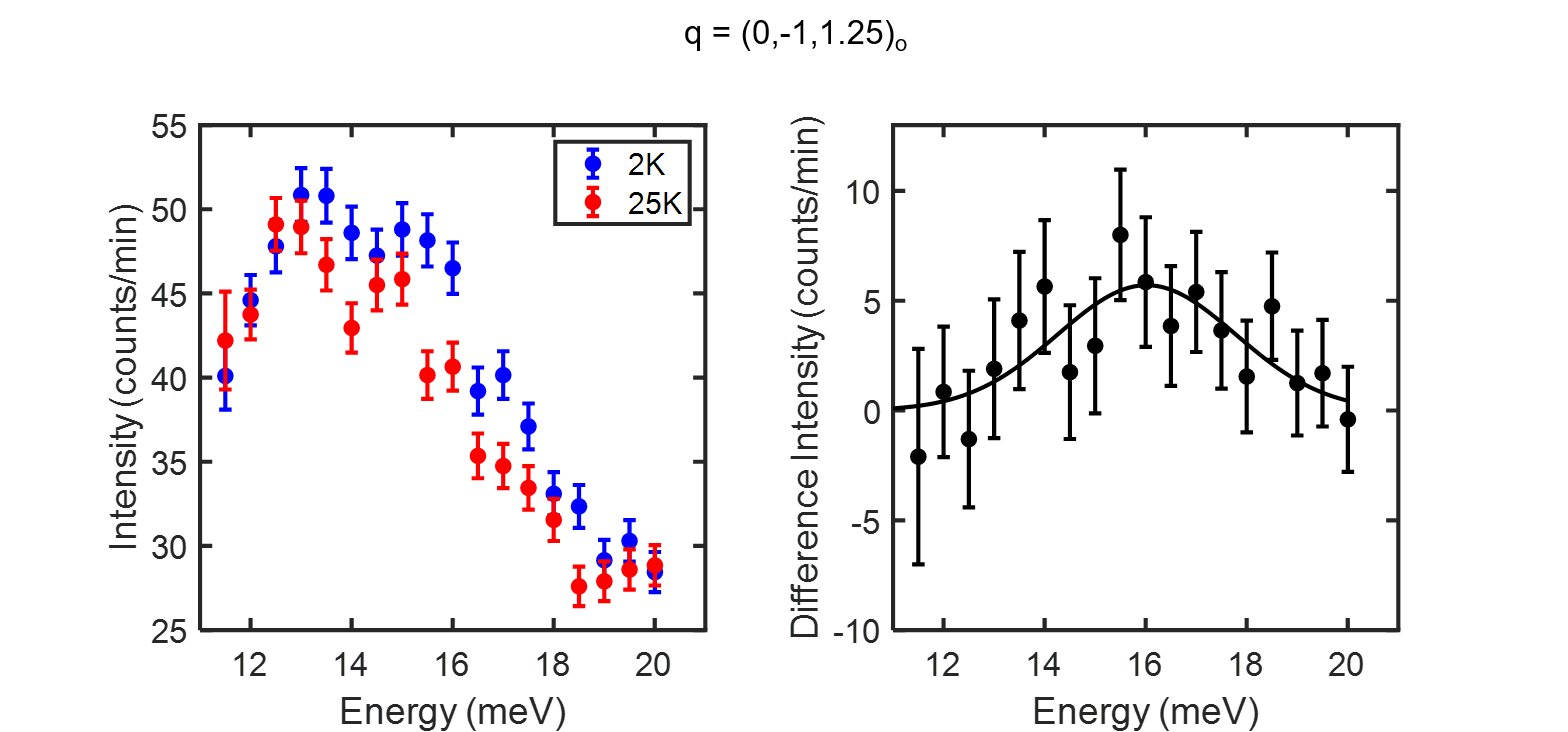}
	\caption{Left: Raw energy scans for $x = 0.05$ and $(0,-1,1.25)_\text{o}$ at 2 K (blue circles) and 25 K (red circles). Right: Difference between the raw scans (black circles) and fit to gaussian profile (solid black line).}
	\label{fig:YCa05_011P25}
\end{figure}

\begin{figure}[!htb]
	\includegraphics[width=0.7\textwidth]{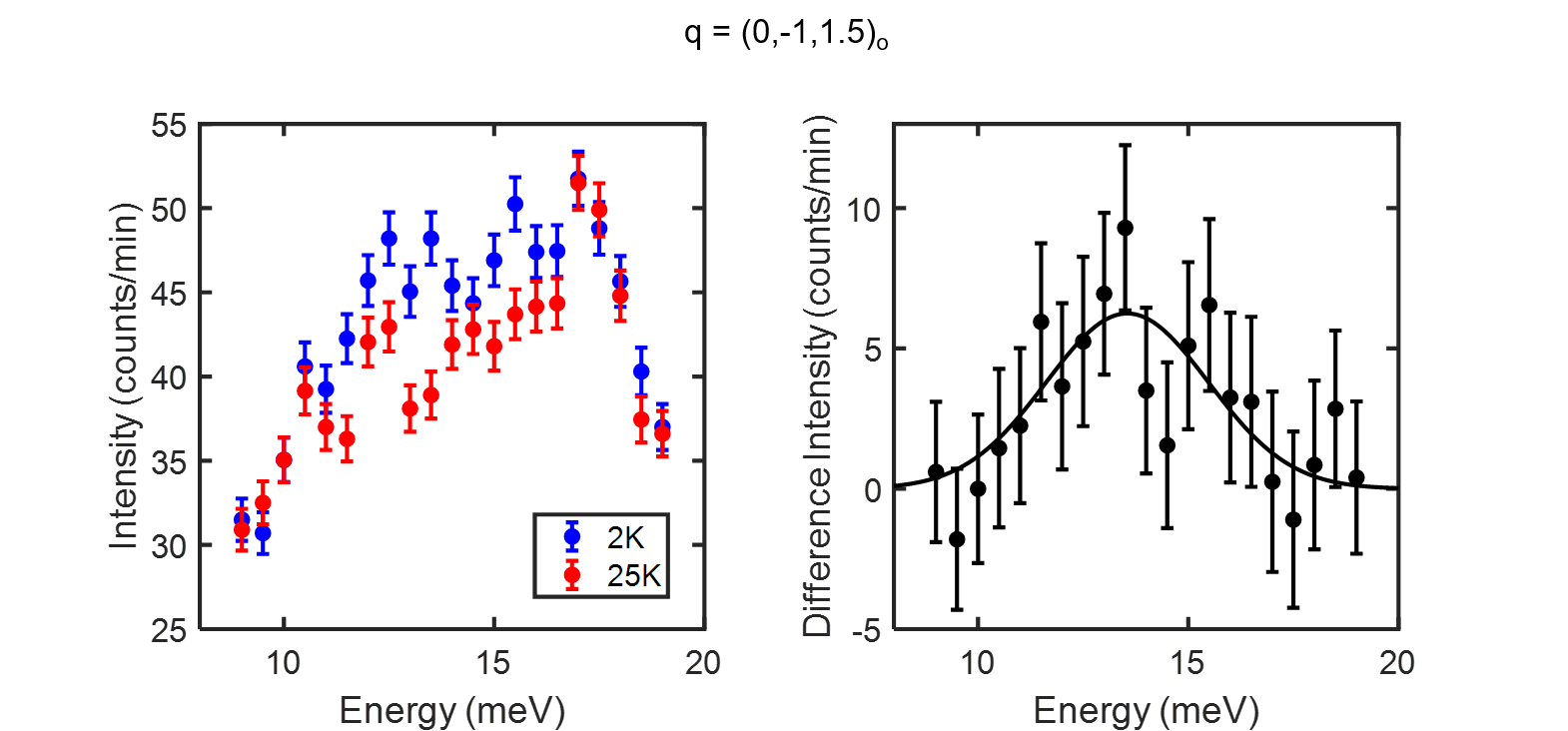}
	\caption{Left: Raw energy scans for $x = 0.05$ and $(0,-1,1.5)_\text{o}$ at 2 K (blue circles) and 25 K (red circles). Right: Difference between the raw scans (black circles) and fit to gaussian profile (solid black line).}
	\label{fig:YCa05_011P5}
\end{figure}

\begin{figure}[!htb]
	\includegraphics[width=0.7\textwidth]{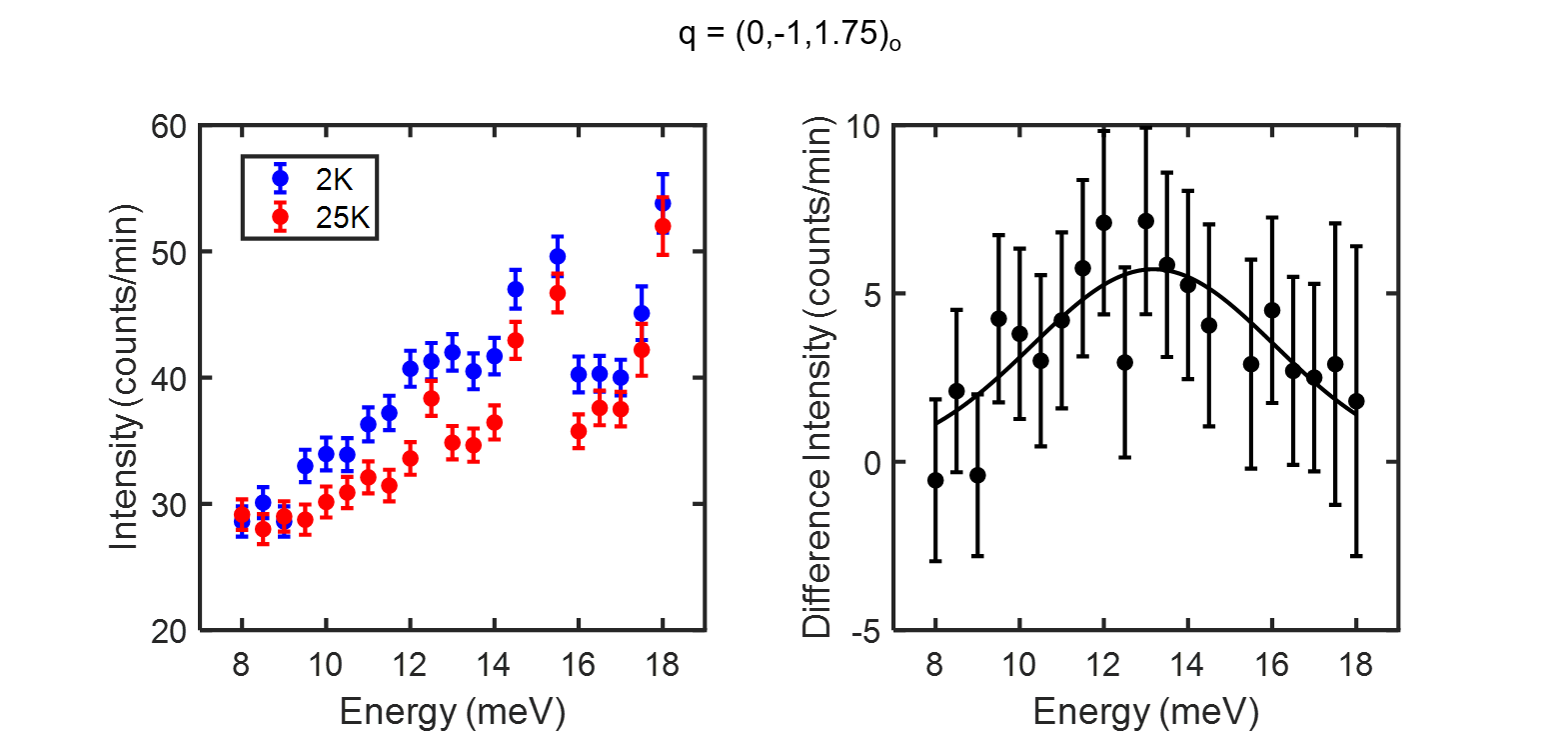}
	\caption{Left: Raw energy scans for $x = 0.05$ and $(0,-1,1.75)_\text{o}$ at 2 K (blue circles) and 25 K (red circles). Right: Difference between the raw scans (black circles) and fit to gaussian profile (solid black line).}
	\label{fig:YCa05_011P75}
\end{figure}

\begin{figure}[!htb]
	\includegraphics[width=0.7\textwidth]{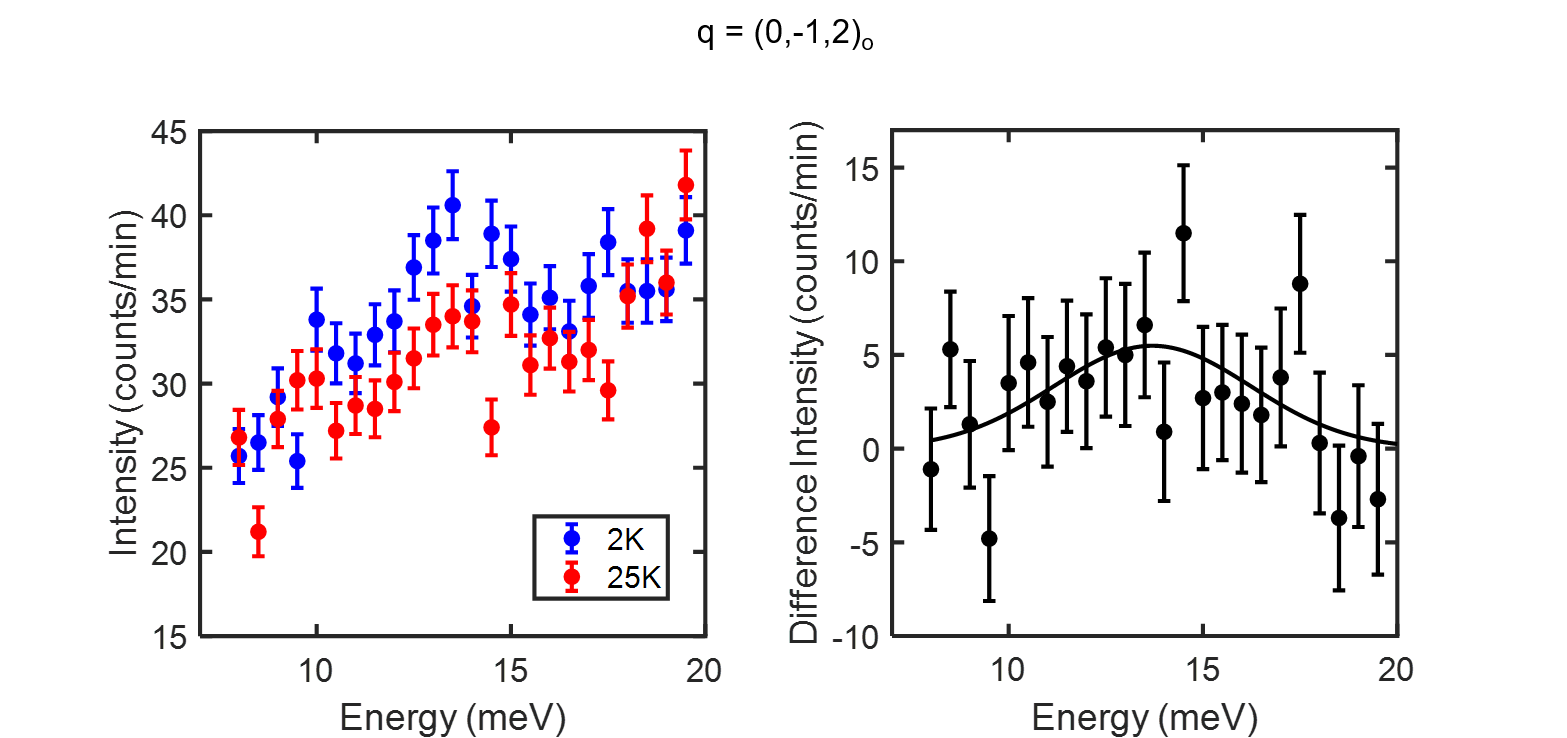}
	\caption{Left: Raw energy scans for $x = 0.05$ and $(0,-1,2)_\text{o}$ at 2 K (blue circles) and 25 K (red circles). Right: Difference between the raw scans (black circles) and fit to gaussian profile (solid black line).}
	\label{fig:YCa05_012}
\end{figure}

\begin{figure}[!htb]
	\includegraphics[width=0.7\textwidth]{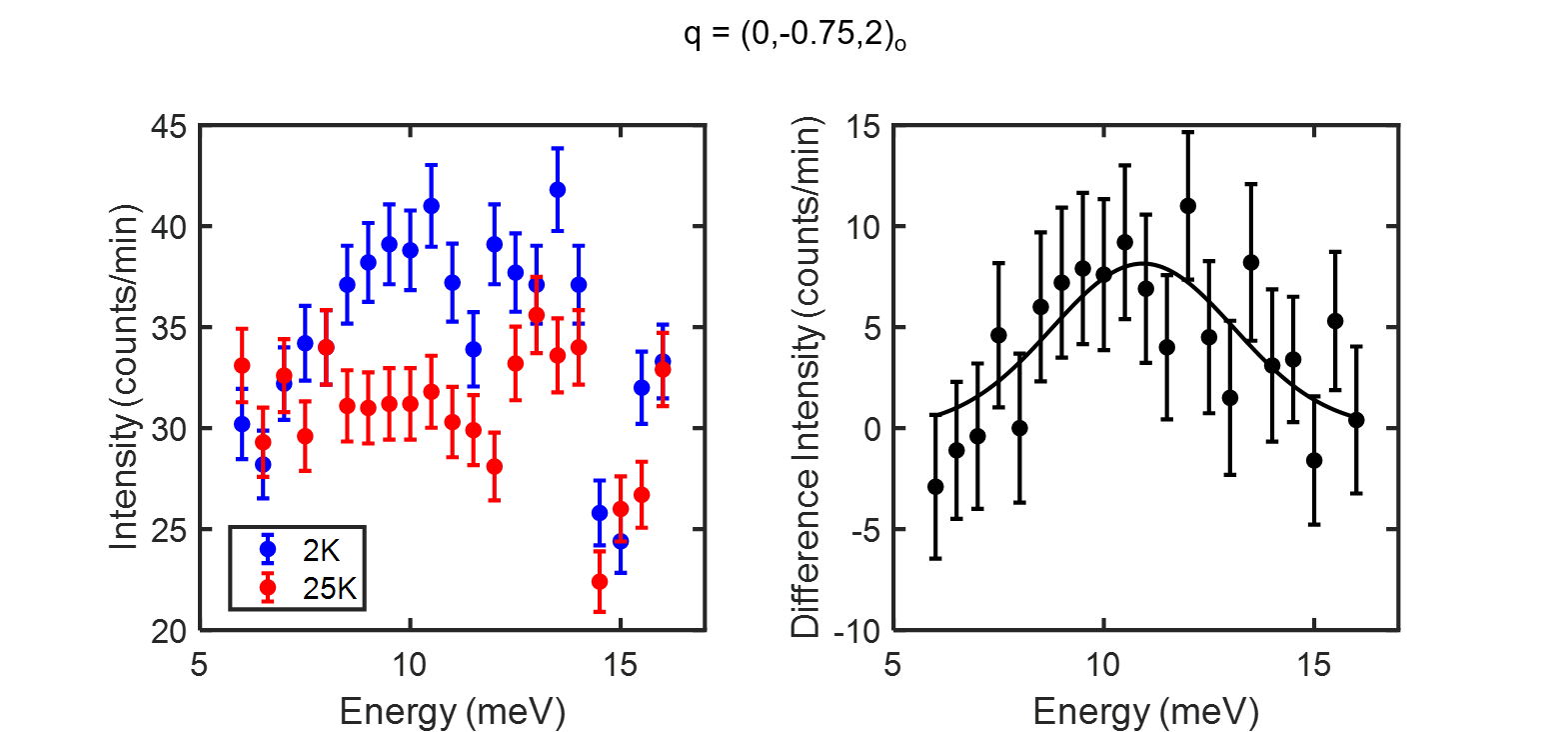}
	\caption{Left: Raw energy scans for $x = 0.05$ and $(0,-0.75,2)_\text{o}$ at 2 K (blue circles) and 25 K (red circles). Right: Difference between the raw scans (black circles) and fit to gaussian profile (solid black line).}
	\label{fig:YCa05_00P752}
\end{figure}

\begin{figure}[!htb]
	\includegraphics[width=0.7\textwidth]{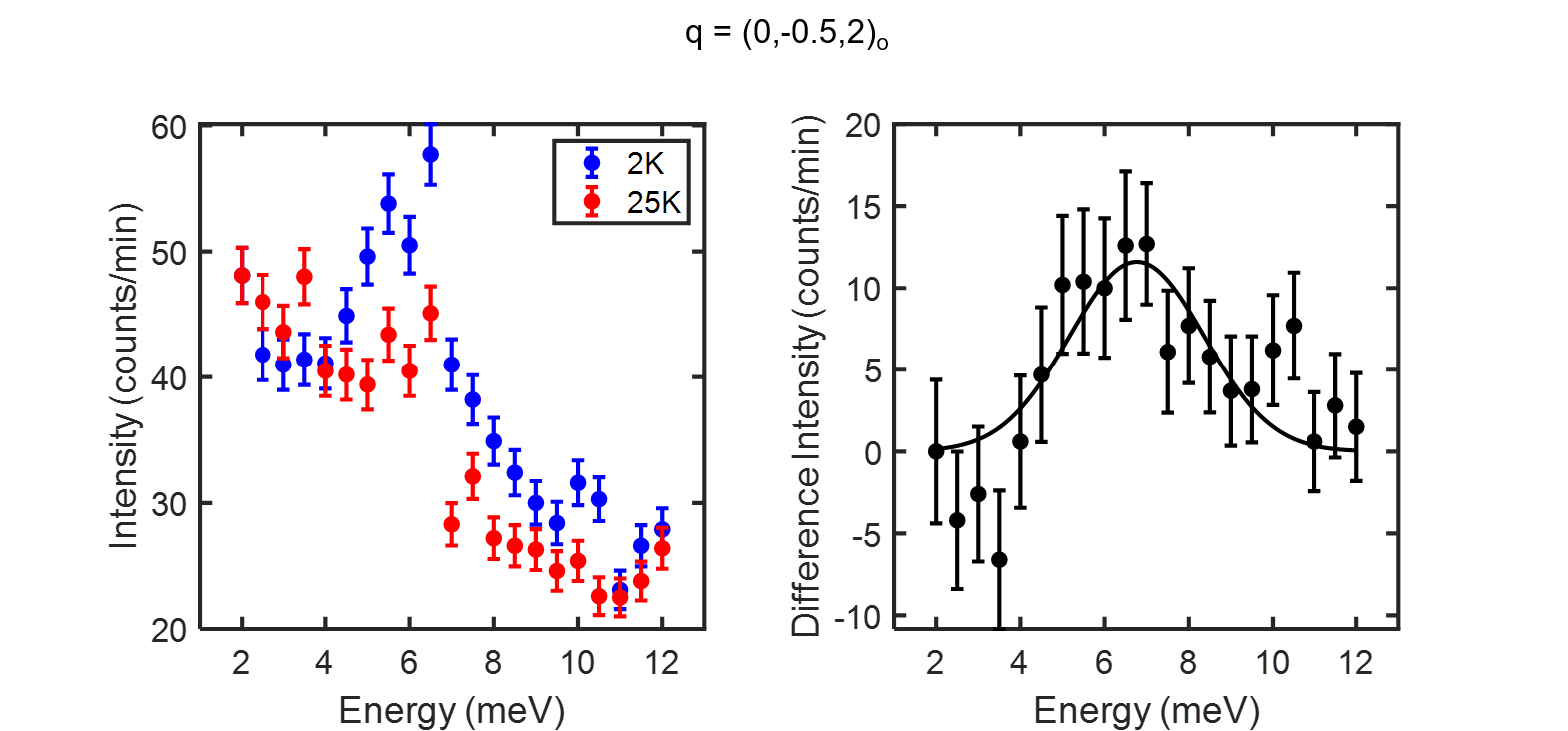}
	\caption{Left: Raw energy scans for $x = 0.05$ and $(0,-0.5,2)_\text{o}$ at 2 K (blue circles) and 25 K (red circles). Right: Difference between the raw scans (black circles) and fit to gaussian profile (solid black line).}
	\label{fig:YCa05_00P52}
\end{figure}

\clearpage
\subsection{$x = 0.15$}

\begin{figure}[!htb]
	\includegraphics[width=0.7\textwidth]{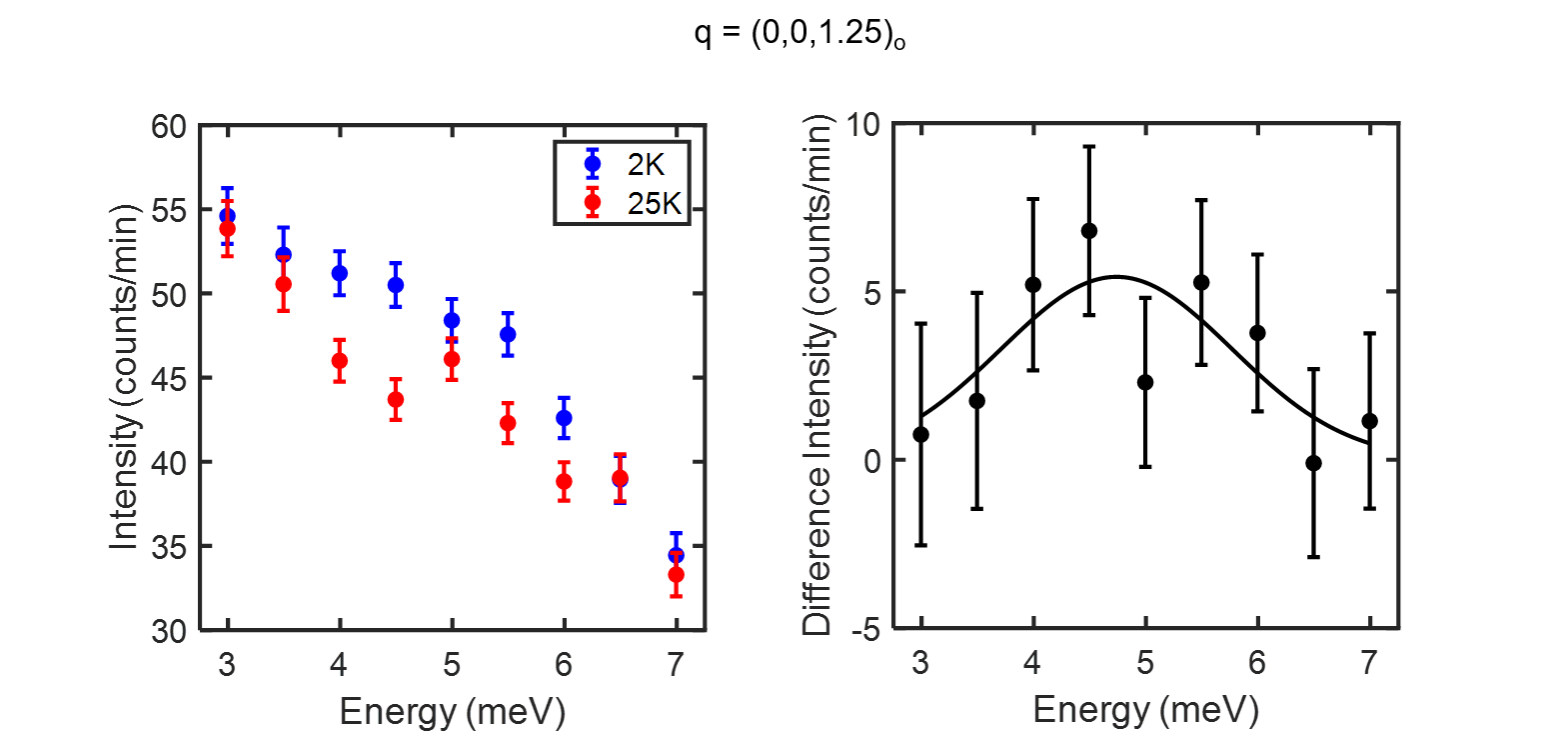}
	\caption{Left: Raw energy scans for $x = 0.10$ and $(0,0,1.25)_\text{o}$ at 2 K (blue circles) and 25 K (red circles). Right: Difference between the raw scans (black circles) and fit to gaussian profile (solid black line).}
	\label{fig:YCa10_001P25}
\end{figure}

\begin{figure}[!htb]
	\includegraphics[width=0.7\textwidth]{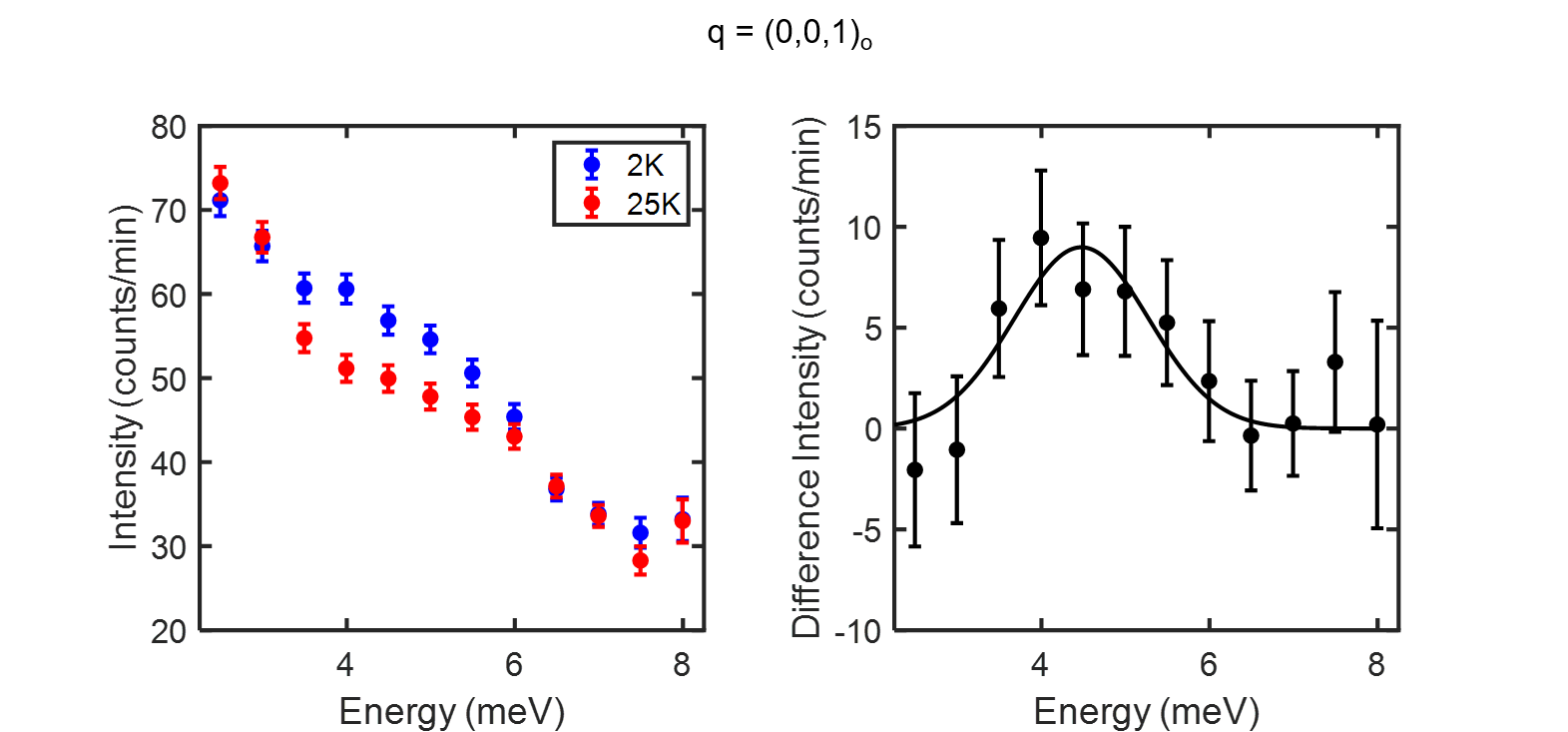}
	\caption{Left: Raw energy scans for $x = 0.10$ and $(0,0,1)_\text{o}$ at 2 K (blue circles) and 25 K (red circles). Right: Difference between the raw scans (black circles) and fit to gaussian profile (solid black line).}
	\label{fig:YCa10_001}
\end{figure}

\begin{figure}[!htb]
	\includegraphics[width=0.7\textwidth]{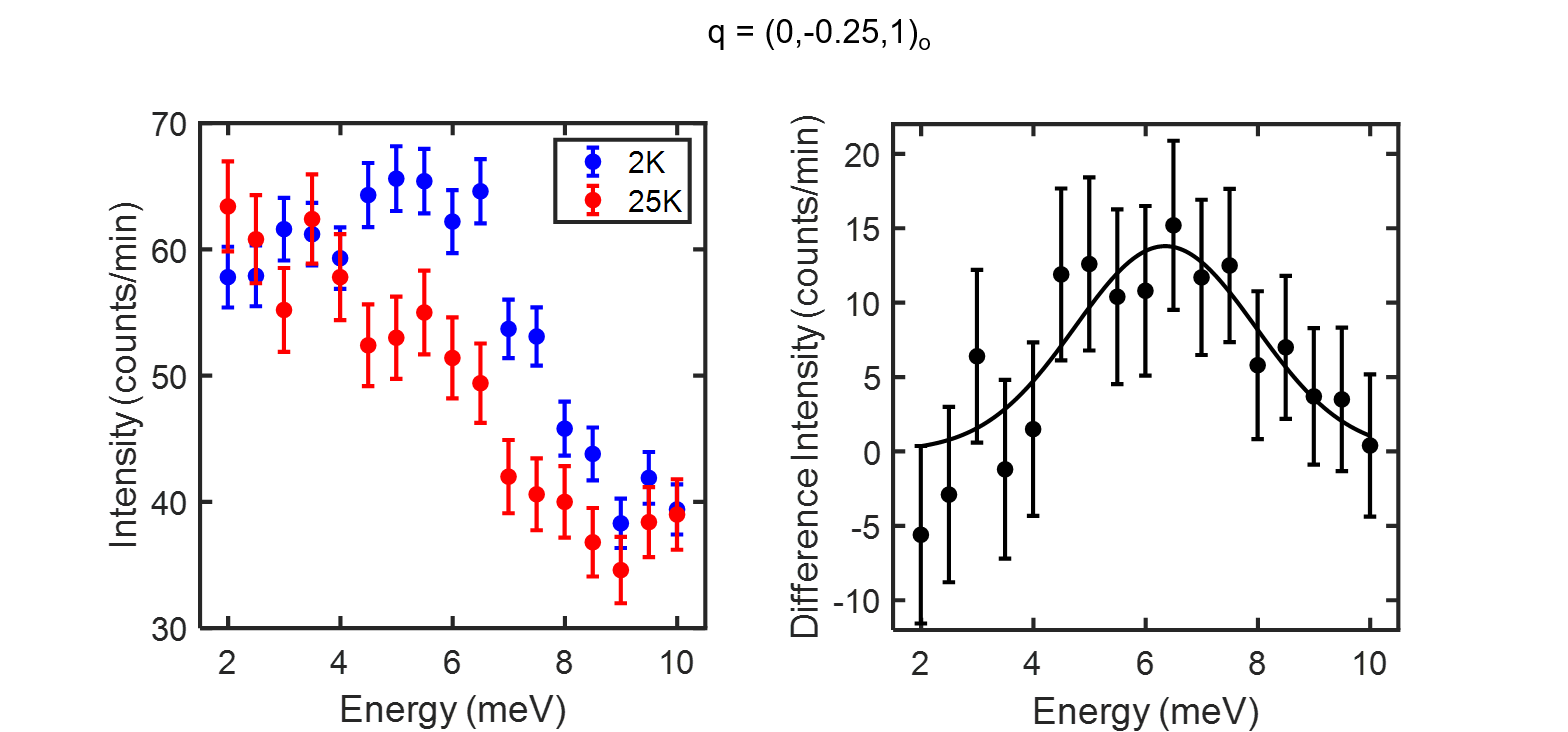}
	\caption{Left: Raw energy scans for $x = 0.10$ and $(0,-0.25,1)_\text{o}$ at 2 K (blue circles) and 25 K (red circles). Right: Difference between the raw scans (black circles) and fit to gaussian profile (solid black line).}
	\label{fig:YCa10_00P251}
\end{figure}

\begin{figure}[!htb]
	\includegraphics[width=0.7\textwidth]{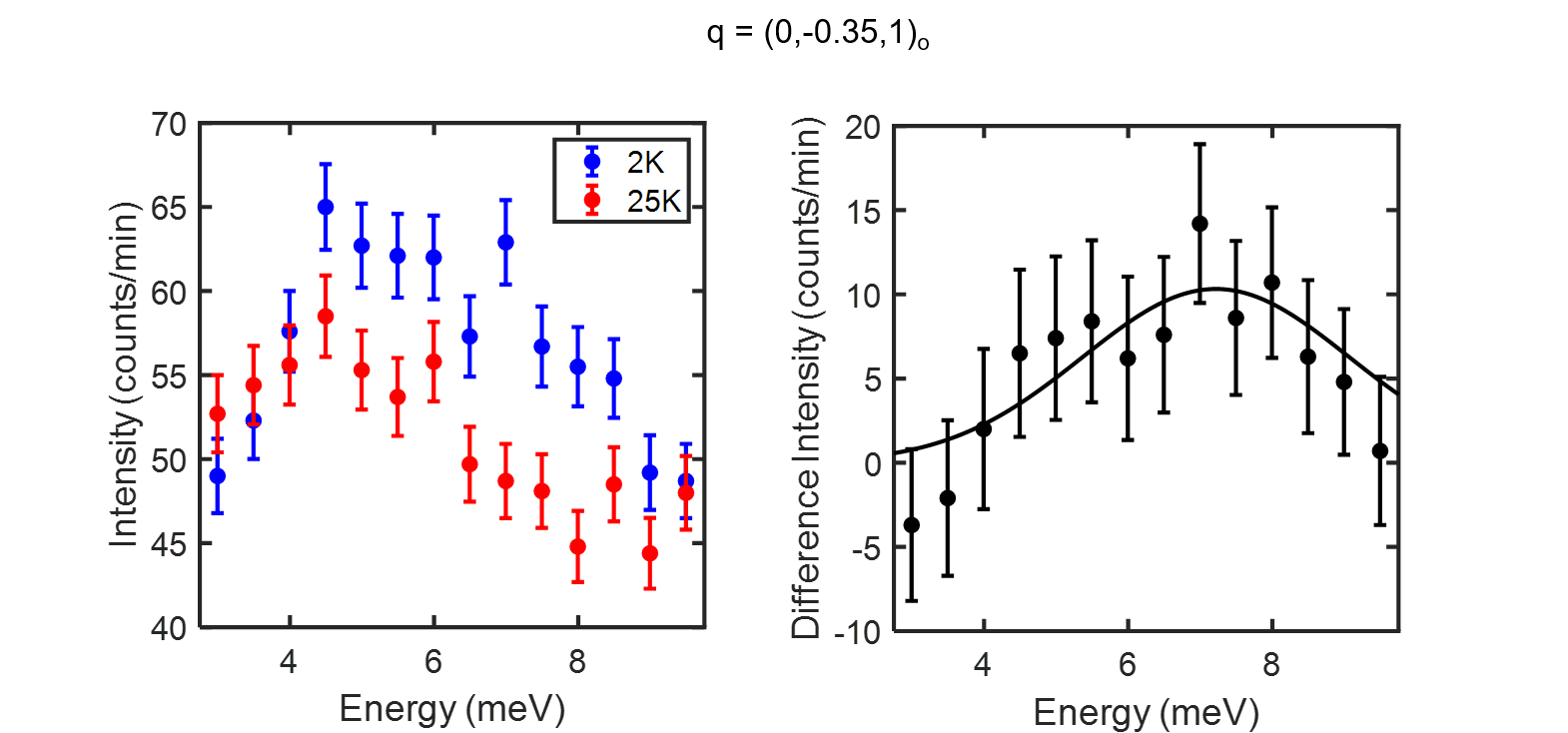}
	\caption{Left: Raw energy scans for $x = 0.10$ and $(0,-0.35,1)_\text{o}$ at 2 K (blue circles) and 25 K (red circles). Right: Difference between the raw scans (black circles) and fit to gaussian profile (solid black line).}
	\label{fig:YCa10_00P351}
\end{figure}

\begin{figure}[!htb]
	\includegraphics[width=0.7\textwidth]{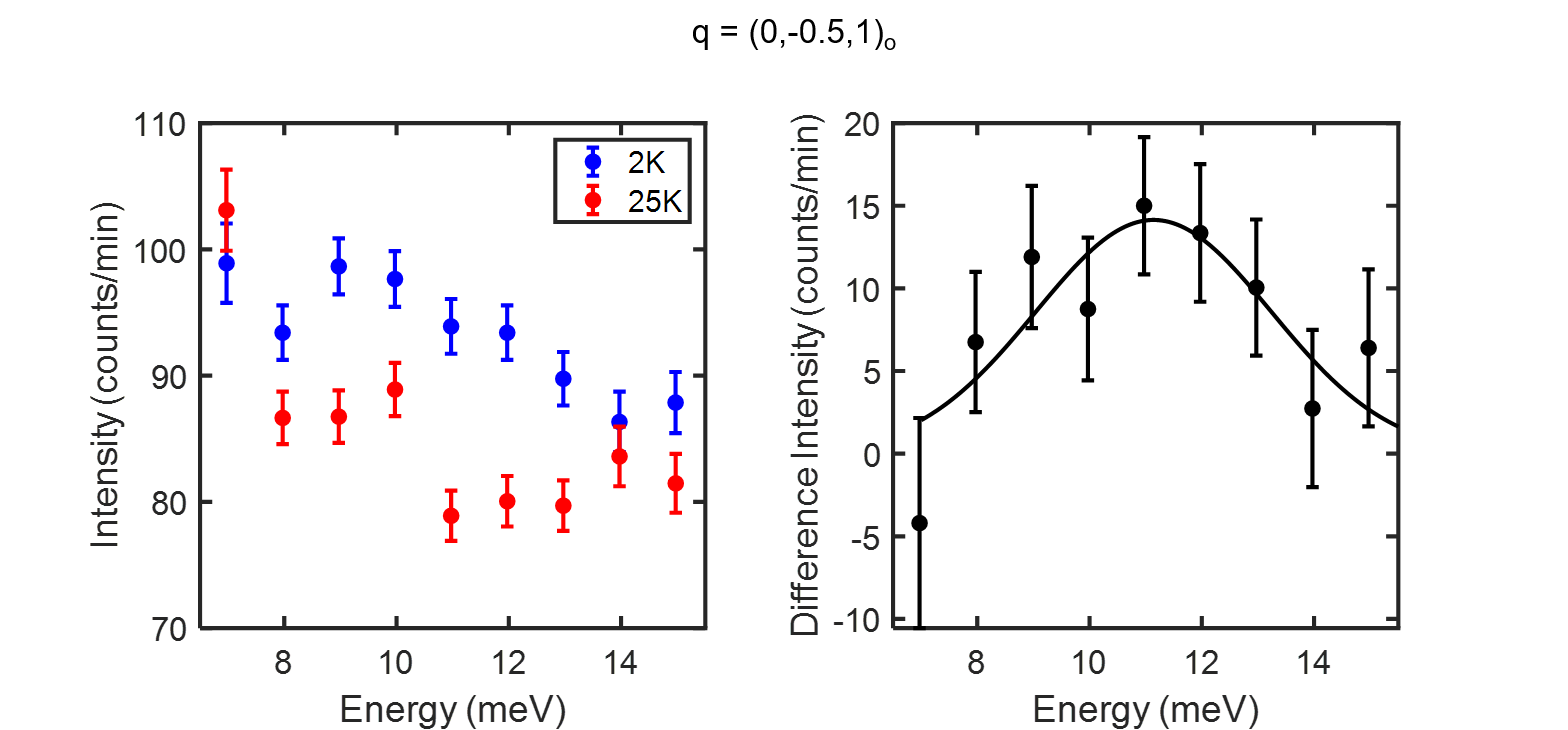}
	\caption{Left: Raw energy scans for $x = 0.10$ and $(0,-0.5,1)_\text{o}$ at 2 K (blue circles) and 25 K (red circles). Right: Difference between the raw scans (black circles) and fit to gaussian profile (solid black line).}
	\label{fig:YCa10_00P51}
\end{figure}

\begin{figure}[!htb]
	\includegraphics[width=0.7\textwidth]{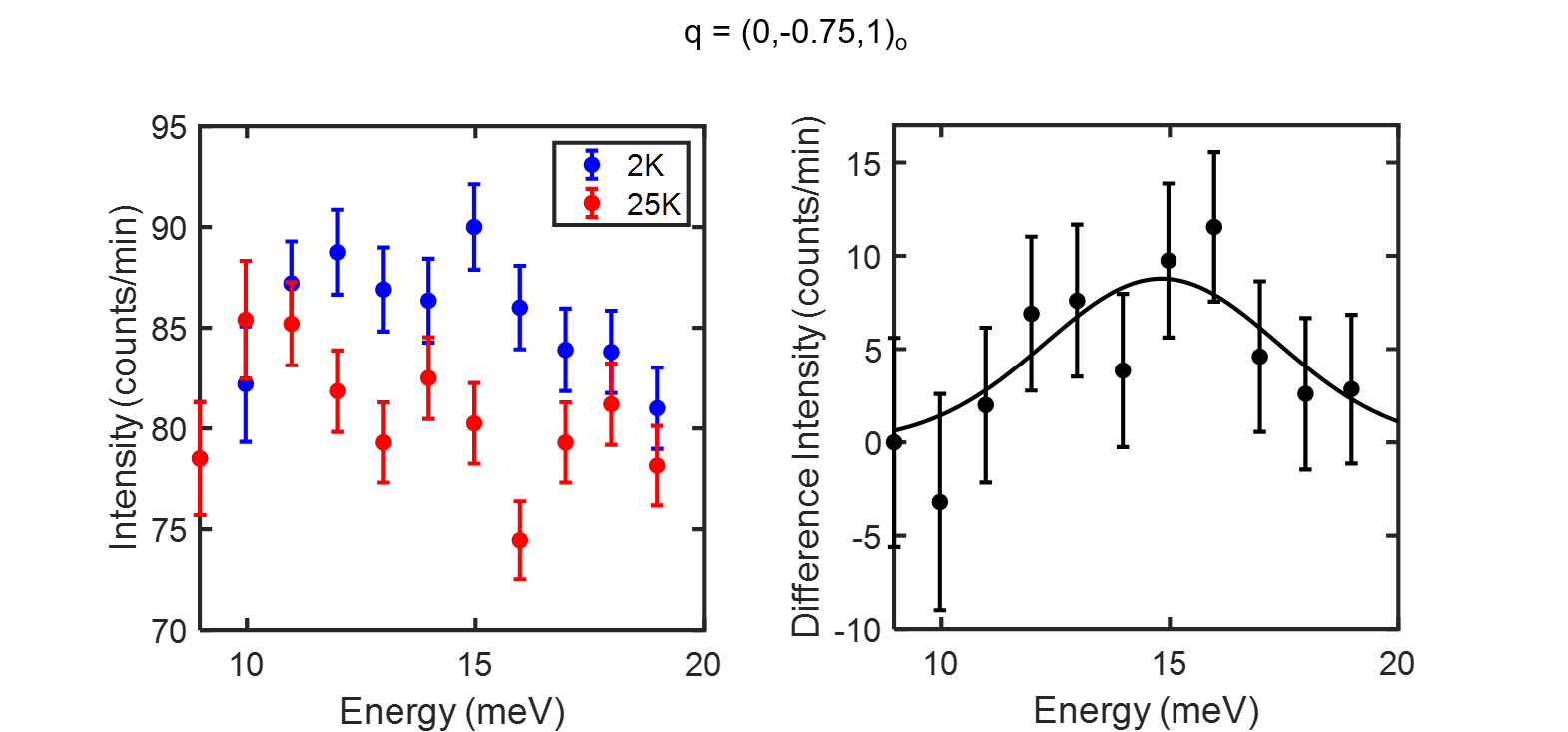}
	\caption{Left: Raw energy scans for $x = 0.10$ and $(0,-0.75,1)_\text{o}$ at 2 K (blue circles) and 25 K (red circles). Right: Difference between the raw scans (black circles) and fit to gaussian profile (solid black line).}
	\label{fig:YCa10_00P751}
\end{figure}

\begin{figure}[!htb]
	\includegraphics[width=0.7\textwidth]{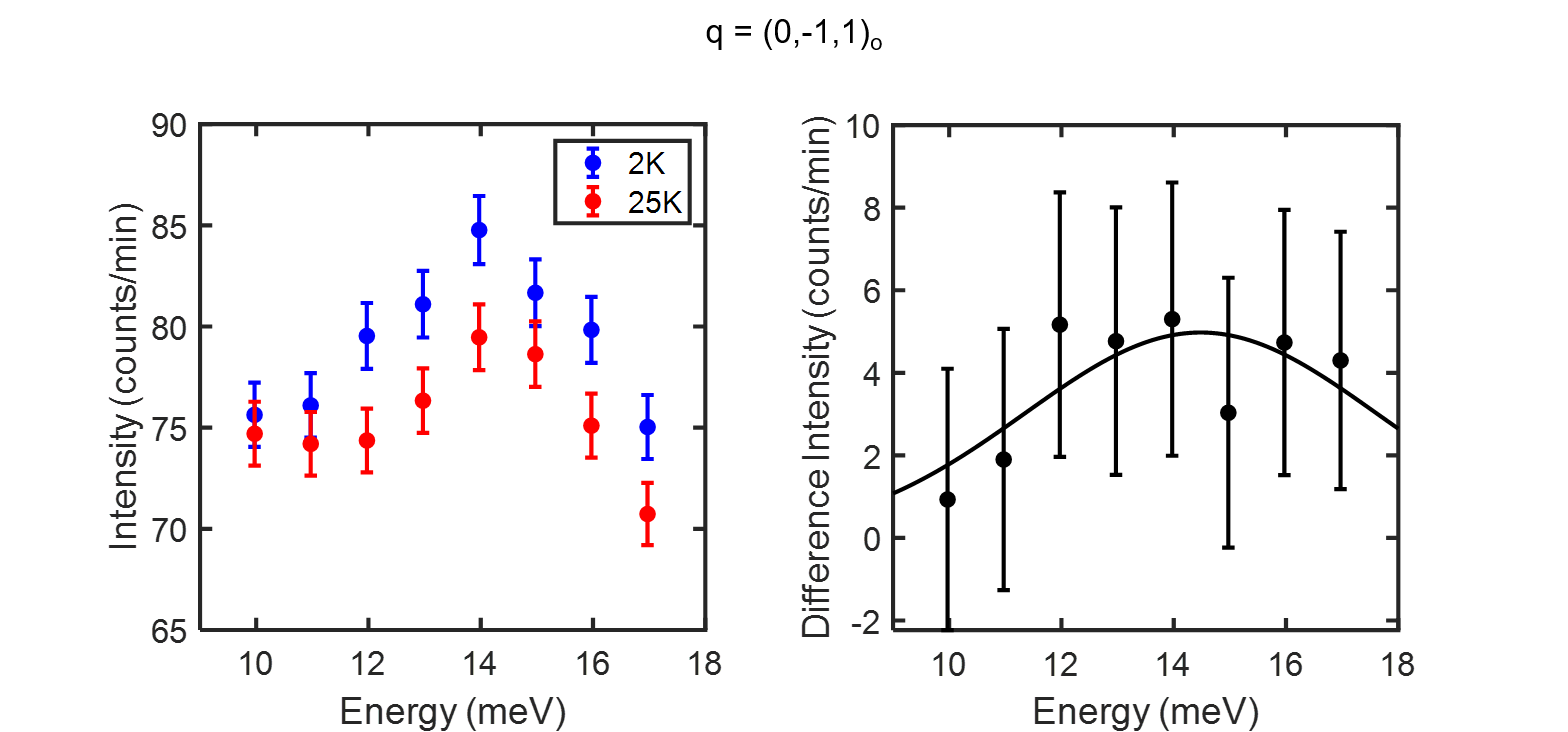}
	\caption{Left: Raw energy scans for $x = 0.10$ and $(0,-1,1)_\text{o}$ at 2 K (blue circles) and 25 K (red circles). Right: Difference between the raw scans (black circles) and fit to gaussian profile (solid black line).}
	\label{fig:YCa10_011}
\end{figure}

\begin{figure}[!htb]
	\includegraphics[width=0.7\textwidth]{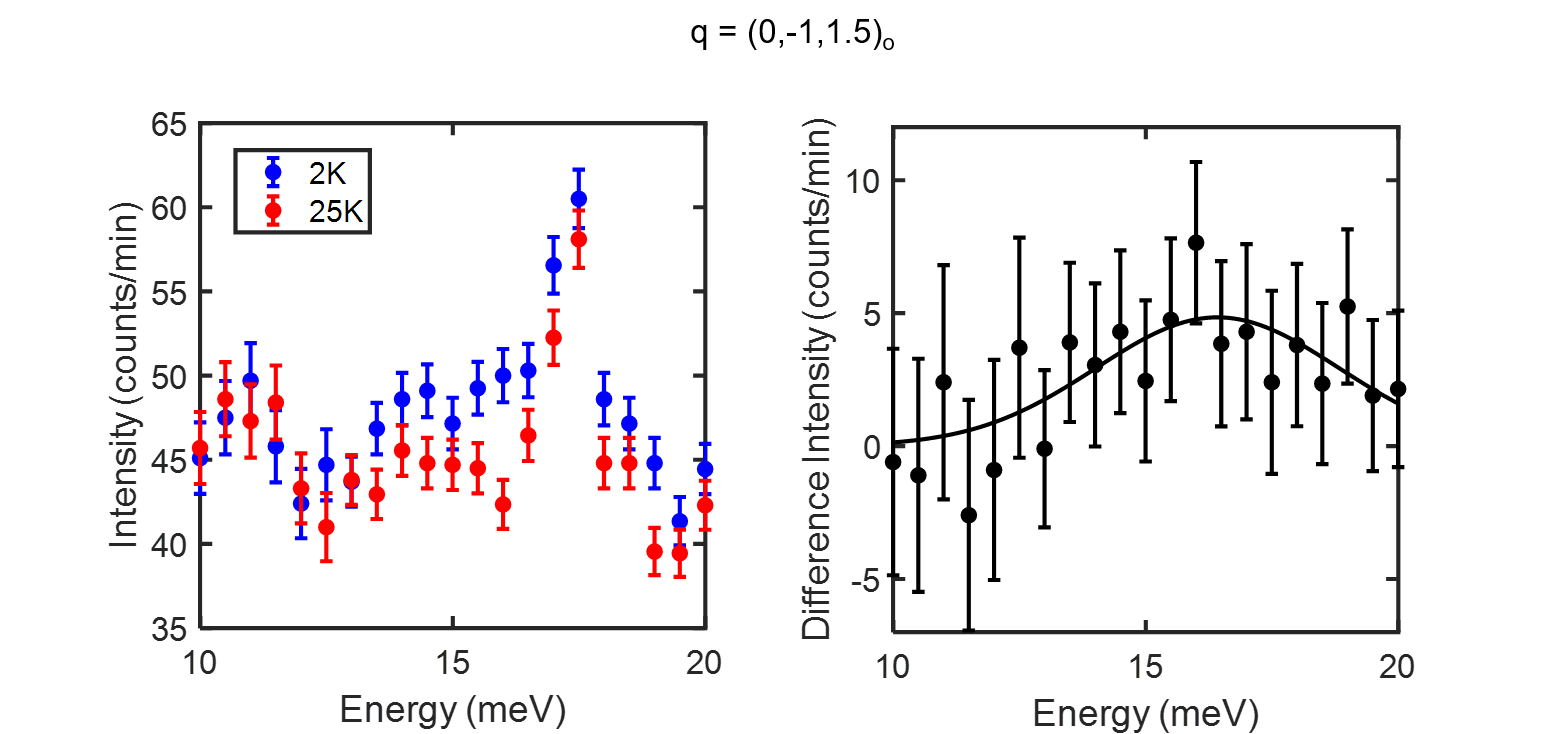}
	\caption{Left: Raw energy scans for $x = 0.10$ and $(0,-1,1.5)_\text{o}$ at 2 K (blue circles) and 25 K (red circles). Right: Difference between the raw scans (black circles) and fit to gaussian profile (solid black line).}
	\label{fig:YCa10_011P5}
\end{figure}

\begin{figure}[!htb]
	\includegraphics[width=0.7\textwidth]{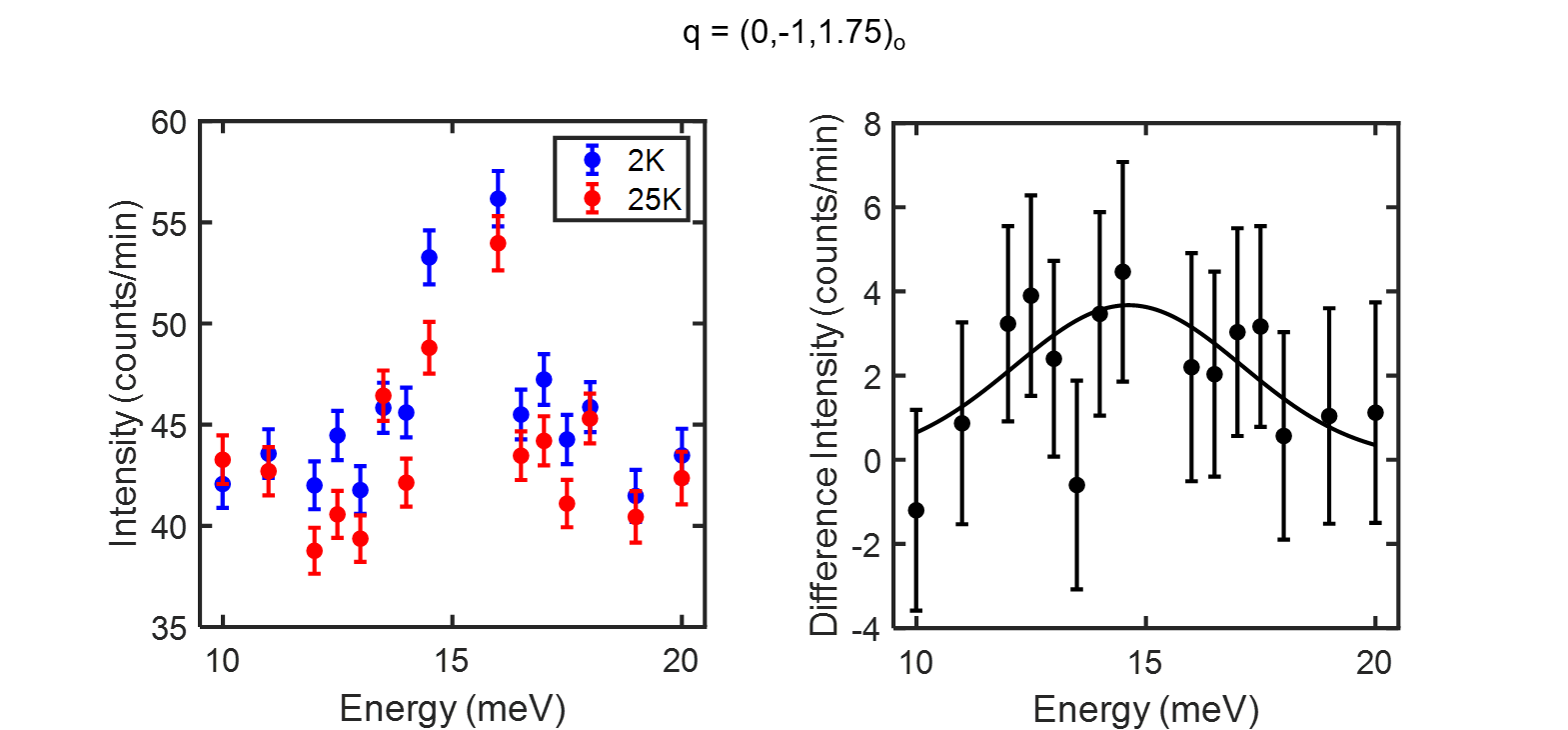}
	\caption{Left: Raw energy scans for $x = 0.10$ and $(0,-1,1.75)_\text{o}$ at 2 K (blue circles) and 25 K (red circles). Right: Difference between the raw scans (black circles) and fit to gaussian profile (solid black line).}
	\label{fig:YCa10_011P75}
\end{figure}

\begin{figure}[!htb]
	\includegraphics[width=0.7\textwidth]{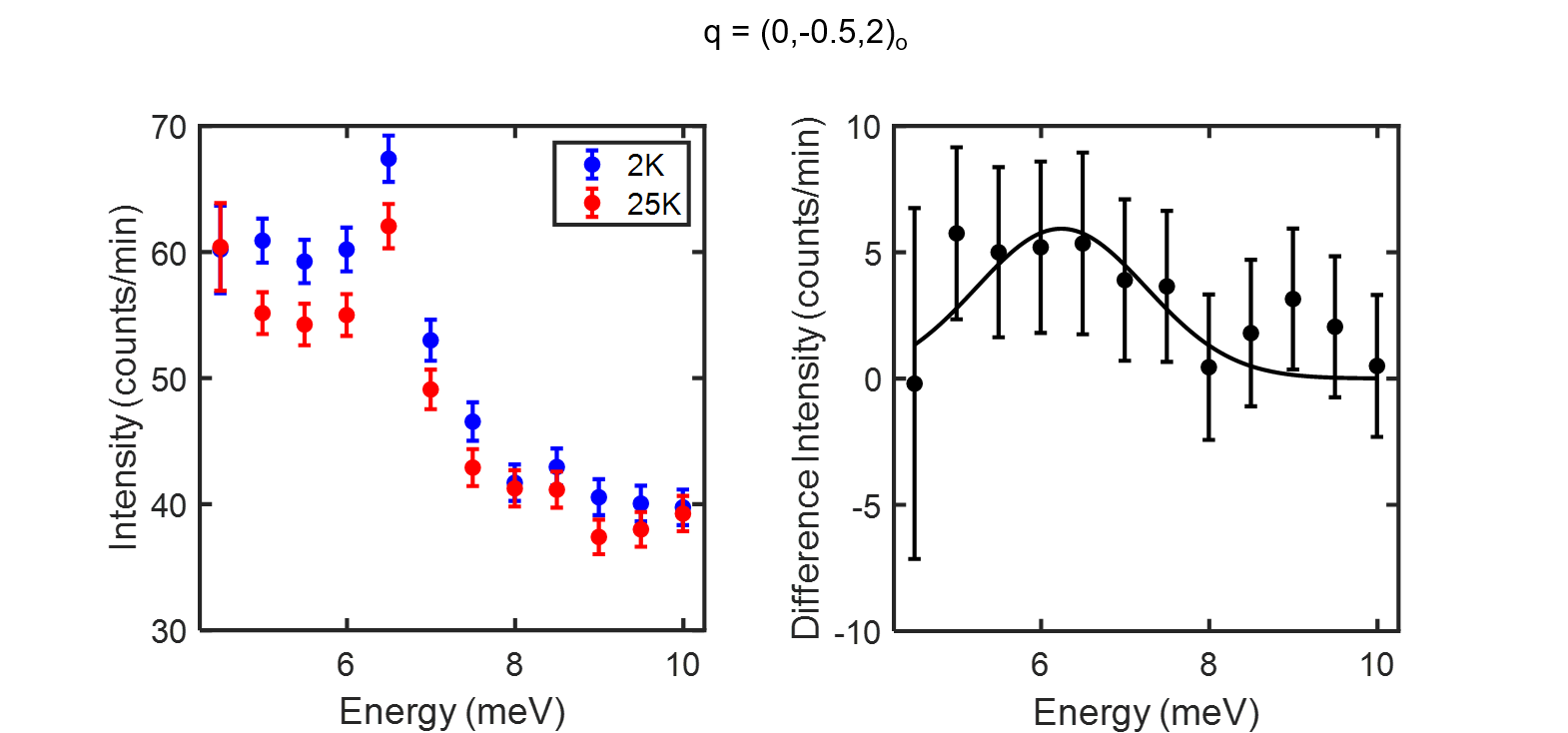}
	\caption{Left: Raw energy scans for $x = 0.10$ and $(0,-0.5,2)_\text{o}$ at 2 K (blue circles) and 25 K (red circles). Right: Difference between the raw scans (black circles) and fit to gaussian profile (solid black line).}
	\label{fig:YCa10_00P52}
\end{figure}

\clearpage

\end{document}